\documentclass[10pt,aps,prc,twocolumn,superscriptaddress,floatfix,nofootinbib]{revtex4-1}
\usepackage{times}
\usepackage[T1]{fontenc}

\usepackage{multirow}

\usepackage{ulem}

\newcommand{\add}[1]{#1}

\usepackage{amsmath,amssymb,mathrsfs}
\usepackage{bm}
\usepackage{ascmac}
\usepackage{geometry}
\usepackage{graphicx}

\def\0\\{\nonumber\\}
\def\bs#1{\boldsymbol{#1}}

\def\zI{\mathrm{i}\hspace{0.2mm}}

\makeatletter
\newcommand\footnoteref[1]{\protected@xdef\@thefnmark{\ref{#1}}\@footnotemark}
\makeatother

\begin{document}

\title{
Time-dependent extension of the self-consistent band theory for neutron star matter:
Anti-entrainment effects in the slab phase
}

\author{Kazuyuki Sekizawa}
\email[]{Present Address: Department of Physics, School of Science, Tokyo Insti-}
\email[]{tute of Technology, Tokyo 152-8551, Japan; sekizawa@phys.titech.ac.jp}
\affiliation{Center for Transdisciplinary Research, Institute for Research Promotion, Niigata University, Niigata 950-2181, Japan}
\affiliation{Nuclear Physics Division, Center for Computational Sciences, University of Tsukuba, Ibaraki 305-8577, Japan}

\author{Sorataka Kobayashi}
\affiliation{Graduate School of Science and Technology, Niigata University, Niigata 950-2181, Japan}

\author{Masayuki Matsuo}
\email[]{matsuo@phys.sc.niigata-u.ac.jp}
\affiliation{Department of Physics, Faculty of Science, Niigata University, Niigata 950-2181, Japan}

\date{April 5, 2022}

\begin{abstract}
\begin{description}
\item[Background]
In the solid crust of neutron stars, a variety of crystalline structure may exist. Recently
the band theory of solids has been applied to the inner crust of neutron stars and
significance of the entrainment between dripped neutrons and the solid crust was advocated.
Since it influences interpretations of various phenomena of neutron stars,
it has been desired to develop deeper understanding of the microphysics behind.

\item[Purpose]
The purpose of the present article is to propose a fully self-consistent microscopic framework
for describing time-dependent dynamics of neutron star matter, which allows us to explore
diverse properties of nuclear matter, including the entrainment effect.

\item[Methods]
A fully self-consistent nuclear band theory is employed with Skyrme SLy4 energy density functional.
A time-dependent extension of the microscopic band theory is developed based on the
time-dependent density functional theory (TDDFT). An intuitive real-time method is proposed
for extracting the collective mass of a nuclear cluster immersed in a sea of dripped neutrons.

\item[Results]
As the first application of the time-dependent self-consistent band theory for nuclear systems,
we investigate the slab phase of nuclear matter with various proton fractions. From a dynamic
response of the system to an external force, we extract the collective mass of a slab, associated
with entrained neutrons as well as bound nucleons. We find that the extracted collective mass is smaller
than a naive estimation based on a potential profile and single-particle energies. We show that
the reduction is mainly caused by ``counterflow'' of dripped neutrons towards the direction opposite
to the motion of the slabs. We interpret it as an ``anti-entrainment'' effect. As a result, the number
of effectively bound neutrons is reduced, indicating an enhancement of the number density of
conduction neutrons. We demonstrate that those findings are consistent with a static treatment
in the band theory of solids.

\item[Conclusions]
The proposed approach offers a new research possibility of investigating non-linear many-body
dynamics of nuclear matter microscopically, taking full account of the periodic structure in
the neutron star crusts. The fully self-consistent band theory calculations, in both static and dynamic
formalisms, suggest that the mobility of dripped neutrons is larger than expected without band
structure effects, at least for the slab phase of nuclear matter.

\end{description}
\end{abstract}


\maketitle

\section{Introduction}

This paper aims at developing a fully microscopic, dynamic approach to investigate
nuclear many-body dynamics in neutron star matter. Especially, we focus on the inner
crust of neutron stars, where dripped (unbound) neutrons permeate a Coulomb lattice
of neutron-rich nuclei. At a bottom layer of the crust, peculiar crystalline nuclear bundles
such as rod- or slab-shaped structure may emerge as a result of competition between
the Coulomb and the nuclear surface energies \cite{Ravenhall(1983),Hashimoto(1984)}.
Such crystalline (periodic) structure effectively couples with dripped neutrons via the Bragg scattering
\cite{Chamel(2017)}, which is called the entrainment effect. Due to the non-dissipative
entrainment, some part of dripped neutrons, which are supposed to be
$^1S_0$ superfluid in cold neutron stars below the critical temperature $T$\,$\lesssim$\,10$^{10}$\,K,
may be immobilized by the periodic structure. A proper quantum mechanical method
to treat a system with a periodic potential is the band theory of solids \cite{Ashcroft-Mermin}.
Although the band theory has been a standard tool in solid-state physics for a long time,
its application to nuclear systems is rather new.

Band structure of dripped neutrons was first investigated in the neutron star context
by Oyamatsu and Yamada in 1994 \cite{Oyamatsu(1994)}, with a simplified one-dimensional
square-well potential. In 2005, Carter et al.~extended the model of Ref.~\cite{Oyamatsu(1994)}
for the slab and rod phases to examine the entrainment effects \cite{Carter(2005)}.
It was shown that the effective mass of conduction neutrons is enhanced by a factor
of 1.02--1.03 and 1.11--1.40 for the slab and rod phases, respectively. Extending and
applying this model, Chamel explored the entrainment effects further in a three-dimensional
(3D) lattice of spherical clusters in low-density region of the inner crust \cite{Chamel(2005),
Chamel(2006),Chamel(2007),Chamel(2012)}. It has been suggested that the neutron
effective mass could be very large, a factor of ten or even more larger than the bare neutron
mass, in some layers of the inner crust. Since such huge effective mass can influence
interpretations of various phenomena of neutron stars, the results of Chamel initiated
debates and have attracted considerable attention.

Since the seminal work by Chamel, many authors considered its impact on neutron
star models. For instance, the entrainment affects long-wavelength collective excitations
of the inner crust such as lattice phonons, Anderson-Bogoliubov modes of neutron superfluid,
and their mutual coupling \cite{Chamel(2013),Kobyakov(2013),Kobyakov(2016),Durel(2018)}.
As a result, heat capacity and thermal conductivity, and thus thermal evolution of the star can
be affected. Besides, the entrainment also affects models of neutron star glitches. In the standard
vortex-mediated glitch scenario \cite{Anderson(1975)}, which explains the quasiperiodic occurrence
and the long post-glitch relaxations, transfer of angular momentum of neutron superfluid to the
solid crust (corotate with the rest of the star) causes a glitch. However, if superfluid neutrons were
strongly entrained by the crust, the inner crust can not store enough angular momentum to explain
large glitches observed in, e.g., the Vela pulser ($\Delta\nu/\nu$\,$\sim$\,10$^{-6}$, where
$\nu$ and $\Delta\nu$ represent the rotational frequency and its change in a glitch event, respectively)
\cite{Andersson(2012),Chamel(2013)glitch}. If it is the case, some part of the core must be
involved in the glitch mechanism (see, e.g., Ref.~\cite{Haskell(2015)} and references therein).
On the other hand, possible avoidance effects have been advocated, such
as neutron superfluidity \cite{Watanabe(2017)} or disorder in crustal structure \cite{Sauls(2020)}.
Besides, apart from the band structure effects, a superfluid hydrodynamic approach indicates that
the coupling between neutron superfluid and the nuclear clusters is rather weak \cite{Martin(2016)}.
Meanwhile, Kashiwaba and Nakatsukasa \cite{Kashiwaba(2019)} reported the first fully self-consistent
band calculations for the slab phase of neutron star matter, based on the nuclear energy density functional
(EDF) approach. They have found that the neutron effective mass is smaller than the bare neutron mass
by a factor of 0.65--0.75 for the slab phase, which is opposite to the effect discussed in Ref.~\cite{Carter(2005)}.
Concerning the importance of the entrainment effects for proper modeling of neutron star physics,
it seems premature to draw a conclusion on the entrainment effects in the inner crust of neutron stars.
In this work, we develop a time-dependent extension of the fully self-consistent nuclear band theory
to shied new light on neutron star physics, including the entrainment effects.

The application of time-dependent mean-field approaches, like time-dependent Hartree-Fock
(TDHF), for nuclear systems already started in the 70s \cite{Negele(review)}. The first application
\cite{BKN(1976)} was actually for collisions of two slab-shaped nuclei, although the analysis at
that time was restricted to isospin symmetric systems, thus without dripped neutrons, with a very
simple interaction, even without the Coulomb force. Since then, numerous studies were performed
for various contexts (see, Refs.~\cite{Nakatsukasa(PTEP),Simenel(review),Nakatsukasa(review),
TDHF-review(2018),Stevenson(2019),Sekizawa(2019)}, for reviews), and its connection to the
time-dependent density functional theory (TDDFT) \cite{DFT1,DFT2,DFT3,DFT4} has been explored
\cite{Nakatsukasa(review)}. It is interesting to note that the application of real-time, real-space
methods, which have been rather common for many years in nuclear physics community, to
atomic/molecular physics was pioneered by Yabana and Bertsch in 1996 \cite{Yabana(1996)},
who experienced in nuclear physics. Later, the method was extended to account for crystalline
solids, i.e.~infinite systems with a periodic potential, by Bertsch et al.~in 2000 \cite{Bertsch(2000)}.
There have been continuous developments and applications of TDDFT, as can be seen for instance
in a well-documented open-source software such as \texttt{Octopus} \cite{Octopus} or \texttt{SALMON}
\cite{SALMON}. Nowadays it is at the forefront of exploring light-matter interactions under extreme
conditions (see, e.g., Refs.~\cite{Yabana(2012),Yamada(2019)} and references therein). Now,
we shall take back the TDDFT approach for crystalline solids to the nuclear physics context,
that is, nuclear dynamics in the inner crust of neutron stars.

In the present paper, as the first step to show the usefulness of the newly developed approach,
we perform TDDFT calculations based on the fully self-consistent band theory for the slab phase
of nuclear matter, neglecting the spin-orbit coupling and pairing correlations. Specifically, we
investigate the collective mass of a slab immersed in a sea of dripped neutrons. We note that
it is not obvious how to define the collective mass of such a quantum object embedded in a Fermi
gas, and it is by itself an academically intriguing problem of its kind. We propose a simple and intuitive
real-time method for extracting the collective mass of a nuclear cluster from a dynamic response
of the system to an external force. We can relate the collective mass to the density of conduction
neutrons which are effectively free from the entrainment effects. It is shown that the collective
mass is reduced and hence the conduction neutron number density is enhanced by what we call
an ``anti-entrainment'' effect, new information which was not clear within the static approach.

The article is organized as follows.
In Sec.~\ref{Sec:Formulation}, we recall the self-consistent band theory for the slab phase
of nuclear matter and present its time-dependent extension. We then introduce a real-time
method to extract the collective mass of a nuclear cluster immersed in a sea of dripped neutrons.
In Sec.~\ref{Sec:Results}, we present the results of numerical simulations for several proton
fractions. We discuss physics behind our numerical results in Sec.~\ref{Sec:Discussion}.
A summary and a perspective are given in Sec.~\ref{Sec:Summary}.

\section{Formulation}\label{Sec:Formulation}

\subsection{The Bloch boundary condition}

In the band theory of solids, the single-particle wave functions are
expressed, according to the Floquet-Bloch theorem, by modulated plane waves:
\begin{equation}
\psi_{\alpha\bs{k}}^{(q)}(\bs{r},\sigma)
= \frac{1}{\sqrt{V}}\,u_{\alpha\bs{k}}^{(q)}(\bs{r},\sigma)\,e^{\zI\bs{k\cdot r}},
\end{equation}
where the spatial, spin, and isospin coordinates are represented by $\bs{r}$, $\sigma$
($=$\,$\uparrow$ or $\downarrow$), and $q$ ($=$\,n or p), respectively, and
$V$ denotes the normalization volume. The subscripts $\alpha$ and
$\bs{k}$ stand for the band index and the Bloch wave vector, respectively. The
single-particle wave functions satisfy the Bloch boundary condition,
\begin{equation}
\psi_{\alpha\bs{k}}^{(q)}(\bs{r}+\bs{T},\sigma)=e^{\zI\bs{k\cdot T}}\psi_{\alpha\bs{k}}^{(q)}(\bs{r},\sigma),
\end{equation}
where $\bs{T}$ is an arbitrary lattice translation vector. It means that the dimensionless
functions $u_{\alpha\bs{k}}^{(q)}(\bs{r},\sigma)$ have the periodicity of the crystal, i.e.,
$u_{\alpha\bs{k}}^{(q)}(\bs{r}+\bs{T},\sigma)=u_{\alpha\bs{k}}^{(q)}(\bs{r},\sigma)$.
In the following, we will refer to the functions $u_{\alpha\bs{k}}^{(q)}(\bs{r},\sigma)$
as Bloch orbitals.

\begin{figure} [t]
\includegraphics[width=7.8cm]{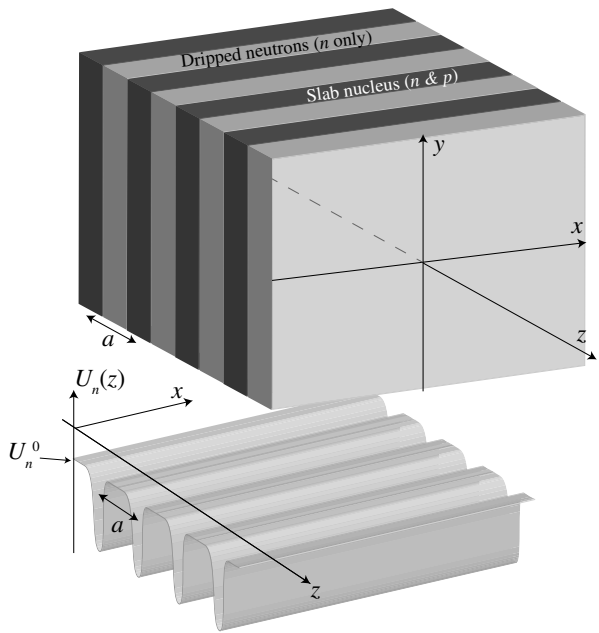}\vspace{2mm}
\caption{
Schematic picture showing the geometry of the systems under study. The nuclear slabs
extend parallel to $xy$ plane and are aligned with a period $a$ along $z$ direction,
as depicted in the upper part of the figure. In the lower part, the nuclear mean-field
potential is depicted in $xz$ plane. Note that when dripped neutrons exist the
maximum value of the mean field potential for neutrons ($U_{\rm n}^0$ in the picture)
is negative, even outside the slabs. The figure has been reprinted from Ref.~\cite{Kashiwaba(2019)}
with permission.
}
\label{Fig:slab}
\end{figure}

In this work, we restrict our analysis to the slab (or ``lasagna'') phase of nuclear
matter at zero temperature, neglecting the spin-orbit coupling and pairing correlations.
We thus omit below the spin index $\sigma$ from the expressions. We consider nuclear
slabs which extend spatially parallel to the $xy$ plane and are in a periodic sequence
along $z$ direction, see Fig.~\ref{Fig:slab}. We assume that the system forms a perfect
crystalline structure of the slabs with a period $a$ (distance between the neighboring slabs).
In this situation, the single-particle wave functions are plane waves along $x$ and $y$ directions
and the lattice vector becomes $\bs{T}=T_x\hat{\bs{e}}_x+T_y\hat{\bs{e}}_y+na\hat{\bs{e}}_z$,
where $T_x$ and $T_y$ are arbitrary real numbers, $n$ is an arbitrary integer, and $\hat
{\bs{e}}_\mu$ is the unit vector along $\mu$ ($=$\,$x$, $y$, or $z$) direction. Then,
the Bloch orbitals depend only on $z$ coordinate,
\begin{equation}
\psi_{\alpha\bs{k}}^{(q)}(\bs{r})
= \frac{1}{\sqrt{V}}\,u_{\alpha\bs{k}}^{(q)}(z)\,e^{\zI\bs{k\cdot r}},
\label{Eq:spwf}
\end{equation}
with the periodicity of the crystal, $u_{\alpha\bs{k}}^{(q)}(z+na)$\,$=$\,$u_{\alpha\bs{k}}^{(q)}(z)$. 

The $z$ component of the Bloch wave vector can be reduced within the first Brillouin
zone, $-\pi/a$\,$\le$\,$k_z$\,$\le$\,$\pi/a$ and, in practice, we discretize it
into $N_{k_z}$ points, i.e. $\Delta k_z$\,$=$\,$\frac{2\pi}{a}\frac{1}{N_{k_z}}$.
This implementation corresponds to a calculation with a length $L\equiv aN_{k_z}$
with the ordinary periodic boundary condition. We normalize the single-particle wave
functions so that the probability of being somewhere within the length $L$ to be unity:
\begin{equation}
\Omega\int_0^L \bigl|\psi_{\alpha\bs{k}}^{(q)}(\bs{r})\bigr|^2{\rm d}z = 1,
\end{equation}
which is equivalent to the normalization condition for the Bloch orbitals,
\begin{equation}
\int_0^a \bigl|u_{\alpha\bs{k}}^{(q)}(z)\bigr|^2\,{\rm d}z \,=\, a.
\label{Eq:normalization}
\end{equation}
Here, $\Omega$ stands for the normalization area such that $V$\,$=$\,$L\Omega$.
In the following sections, we formulate the fully self-consistent band theory for the slab
phase of nuclear matter based on a Skyrme-type EDF.

\subsection{Energy density functionals}\label{Sec:EDF}

We denote the total number of neutrons ($q$\,$=$\,${\rm n}$) or protons ($q$\,$=$\,${\rm p}$)
per unit area within a single period $a$ as $N_q=\int_0^an_q(z)\,{\rm d}z$, where
$n_{\rm n}(z)$ and $n_{\rm p}(z)$ represent neutron and proton number densities,
respectively. The average number density of neutrons ($q$\,$=$\,${\rm n}$) or protons
($q$\,$=$\,${\rm p}$) is then defined as $\bar{n}_q$\,$=$\,$N_q/a$. We also define the
corresponding quantities for baryons (nucleons) as $N_{\rm b}$\,$=$\,$N_{\rm n}+N_{\rm p}$
and $\bar{n}_{\rm b}$\,$=$\,$\bar{n}_{\rm n}+\bar{n}_{\rm p}$. The proton fraction
is defined by $Y_{\rm p}$\,$=$\,$\bar{n}_{\rm p}/\bar{n}_{\rm b}$.

In this work, we adopt the Skyrme SLy4 EDF \cite{Chabanat(1998)} which
has been widely used in various contexts \cite{Bender(2003)}, neglecting
the spin-orbit coupling. The total energy per nucleon reads
\begin{equation}
\frac{E_{\rm nucl}}{A} = \frac{1}{N_{\rm b}}\int_0^a
\bigl( \mathcal{E}_{\rm kin}(z) + \mathcal{E}_{\rm Sky}(z)  + \mathcal{E}_{\rm Coul}^{({\rm p})}(z)\bigr)\,{\rm d}z,
\end{equation}
where the kinetic and the nuclear energy densities are, respectively, given by:
\begin{eqnarray}
\mathcal{E}_{\rm kin}(z)
&=&
\sum_{q={\rm n},{\rm p}}\frac{\hbar^2}{2m_q}\tau_q(z),\\[1mm]
\mathcal{E}_{\rm Sky}(z)
&=& \sum_{t=0,1}\Bigl[
C_t^\rho[n_0]\,n_t^2(z)
+C_t^{\triangle\rho}n_t(z) \partial_z^2 n_t(z)\nonumber\\[-1.5mm]
&&\hspace{6.4mm}
+\;C_t^\tau\bigl(n_t(z)\tau_t(z)-\bs{j}_t^2(z)\bigr)
\Bigr].
\end{eqnarray}
We note that, reflecting an exploratory character of the present study, we will
distinguish neutron and proton masses, which allows us a detailed investigation of
the collective masses as discussed in Sec.~\ref{Sec:Results} and Sec.~\ref{Sec:Discussion}
\footnote{Here, $m_{\rm n}$=939.565\,MeV/$c^2$ and $m_{\rm p}$=938.272\,MeV/$c^2$ are used.}.
Here and henceforth, the symbol $\partial_z$ represents a spatial derivative with
respect to $z$ coordinate. $n_t(z)$, $\tau_t(z)$, and $\bs{j}_t(z)$ denote the
number, kinetic, and momentum densities, respectively. The time-odd momentum
density vanishes in static calculations, while they are, in general, finite in a dynamic
situation. The subscript $t$ specifies isoscalar ($t$\,$=$\,0) and isovector ($t$\,$=$\,1)
densities, which are defined, for the number density for instance, by
$n_0(z)$\,$=$\,$n_{\rm n}(z)+n_{\rm p}(z)$ and $n_1(z)$\,$=$\,$n_{\rm n}(z)-n_{\rm p}(z)$,
respectively (The subscript 0 is often omitted). The detailed expressions of the coefficients
by means of Skyrme force parameters can be found in, e.g., Eq.~(A1) in Ref.~\cite{Lesinski(2007)}
(where symbols like $A_t^{\rm X}$ were used instead of $C_t^{\rm X}$ here). It is to be
reminded that $C_t^\rho[n_0]$ depends on the local number density as $C_t^\rho[n_0]
$\,$=$\,$C_t^\rho+C_{t,{\rm D}}^\rho n_0^\gamma(z)$ as in Ref.~\cite{Kortelainen(2010)}
\footnote{Note that the notation $\gamma$ is used to represent a fractional power
of the isoscalar density in the density-dependent term in Skyrme-type functionals
to avoid confusion with the band index $\alpha$.}.

The neutron and proton densities are formally defined as:
\begin{eqnarray}
n_q(z) &=& 2\sum_{\alpha,\bs{k}}^{\rm occ.}
\bigl|\psi_{\alpha\bs{k}}^{(q)}(\bs{r})\bigr|^2
\label{Eq:def_n_q}\\[1.5mm]
\tau_q(z) &=&
2\sum_{\alpha,\bs{k}}^{\rm occ.}
\bigl|\bs{\nabla}\psi_{\alpha\bs{k}}^{(q)}(\bs{r})\bigr|^2
\label{Eq:def_tau_q}\\[1.5mm]
\bs{j}_q(z) &=&
2\sum_{\alpha,\bs{k}}^{\rm occ.}
{\rm Im}\bigl[\psi_{\alpha\bs{k}}^{(q)*}(\bs{r})\bs{\nabla}\psi_{\alpha\bs{k}}^{(q)}(\bs{r})\bigr].
\label{Eq:def_j_q}
\end{eqnarray}
The summation is taken over all occupied orbitals below the Fermi energy
with the factor of 2 accounting for the spin degeneracy. Note that, in general,
the each component ($i=x,y,z$) of the Bloch wave vector $\bs{k}$ is discretized
to be $k_i=\frac{2\pi}{L_i}n_i$ ($n_i=0,\pm1,\dots$) for an arbitrary system with
a unit cell with length $L_i$ for each direction, and thus $\sum_{\bs{k}}$ should be
regarded as a summation over those integers $n_i$. For the slab phase, the summations
for $k_x$ and $k_y$ can be replaced with integrals,
\begin{equation}
\sum_{k_x,k_y} \rightarrow
\int\frac{\Omega\,{\rm d}k_x{\rm d}k_y}{(2\pi)^2} = \int\frac{\Omega\,k_\parallel{\rm d}k_\parallel}{2\pi},
\label{Eq:sum2int_k_parallel}
\end{equation}
where $k_\parallel$ represents the absolute value of the Bloch wave vector parallel
to the slabs, $k_\parallel\equiv(k_x^2+k_y^2)^{1/2}$. By substituting the Bloch
wave function \eqref{Eq:spwf} into Eqs.~\eqref{Eq:def_n_q}--\eqref{Eq:def_j_q}
with the replacement \eqref{Eq:sum2int_k_parallel}, the densities can be expressed
in terms of the Bloch orbitals as follows:
\begin{eqnarray}
n_q(z) &=&
\frac{1}{\pi L}\sum_{\alpha,k_z} \int k_\parallel\,
\bigl|u_{\alpha\bs{k}}^{(q)}(z)\bigr|^2\,
\theta\bigl(\mu_q - \varepsilon_{\alpha\bs{k}}^{(q)}\bigr)\,{\rm d}k_\parallel,
\label{Eq:n_q}\\[1.5mm]
\tau_q(z) &=&
\frac{1}{\pi L}\sum_{\alpha,k_z} \int k_\parallel\, \Bigl[
k_\parallel^2\, \bigl| u_{\alpha\bs{k}}^{(q)}(z) \bigr|^2
+ \bigl| (\partial_z+\zI k_z) u_{\alpha\bs{k}}^{(q)}(z) \bigr|^2 \Bigr] \nonumber\\[-2.5mm]
&&\hspace{15mm}\times\,\theta\bigl(\mu_q - \varepsilon_{\alpha\bs{k}}^{(q)}\bigr)\,{\rm d}k_\parallel,
\label{Eq:tau_q}\\[1.5mm]
\bs{j}_q(z) &=&
\frac{1}{\pi L}\sum_{\alpha,k_z}\int k_\parallel\,
{\rm Im}\Bigl[u_{\alpha\bs{k}}^{(q)*}(z) (\partial_z+\zI k_z) u_{\alpha\bs{k}}^{(q)}(z)\Bigr]\hat{\bs{e}}_z  \nonumber\\[-2.5mm]
&&\hspace{15mm}\times\,\theta\bigl(\mu_q - \varepsilon_{\alpha\bs{k}}^{(q)}\bigr)\,{\rm d}k_\parallel,
\label{Eq:j_q}
\end{eqnarray}
where $\mu_q$ denotes the chemical potential (Fermi energy), and
$\theta(x)$ is the step function which is one for $x$\,$>$\,0 and zero
otherwise. For the slab phase, $x$ and $y$ components of the momentum
density naturally vanish, i.e. $\bs{j}_q(z)$\,$=$\,$j_{z,q}(z)\hat{\bs{e}}_z$,
assuming that positive and negative wave-number states are equally occupied.

The proton part of the Coulomb EDF is given by
\begin{eqnarray}
\mathcal{E}_{\rm Coul}^{({\rm p})}(z)
&=& \frac{1}{2}V_{\rm Coul}(z)n_{\rm p}(z)
-\frac{3e^2}{4}\biggl(\frac{3}{\pi}\biggr)^{1/3}n_{\rm p}^{4/3}(z),\;\;
\label{Eq:E_Coul_proton}
\end{eqnarray}
where $e$ denotes the elementary charge. The first term corresponds to the
direct term, while the second term corresponds to the exchange term with the
Slater approximation. The Coulomb potential for protons $V_{\rm Coul}$
($-V_{\rm Coul}$ for electrons) is determined by solving the Poisson equation,
\begin{equation}
\frac{{\rm d}^2}{{\rm d}z^2}V_{\rm Coul}(z) = -\frac{e^2}{\varepsilon_0}n_{\rm ch}(z),
\end{equation}
where $\varepsilon_0$ is the vacuum permittivity. 
$n_{\rm ch}(z)$\,$\equiv$\,$n_{\rm p}(z)-n_{\rm e}$ denotes the charge
density, neglecting the charge form factor of protons. Electrons are assumed to be uniformly
distributed with the density $n_{\rm e}$\,$=$\,$\bar{n}_{\rm p}$. The Coulomb potential
is subjected to the charge neutrality condition, $\int_0^a V_{\rm Coul}(z)\,{\rm d}z$\,$=$\,0.

For the electrons' EDF we follow the same procedure as in Ref.~\cite{Kashiwaba(2019)},
where electrons are treated as uniform, degenerated relativistic Fermi gas. The energy of
electrons divided by the baryon (nucleon) number $A$ is given by
\begin{equation}
\frac{E_{\rm elec}}{A} = \frac{1}{N_{\rm b}}\bigl( \mathcal{E}_{\rm kin}^{({\rm e})} + \mathcal{E}_{\rm Coul}^{({\rm e})} \bigr)\,a,
\end{equation}
where the kinetic and the Coulomb energy densities are, respectively, read
\begin{eqnarray}
\mathcal{E}_{\rm kin}^{({\rm e})} &=& \int_0^{p_{\rm F}}\frac{4\pi p^2dp}{(2\pi)^3}\sqrt{m_{\rm e}^2c^4+p^2c^2} \nonumber\\[1mm]
&=& \frac{m_{\rm e}^4c^5}{32\pi^2\hbar^3}(\sinh4\theta_{\rm F}-4\theta_{\rm F}), \\[2mm]
\mathcal{E}_{\rm Coul}^{({\rm e})} &=& -\frac{3e^2}{4}\biggl(\frac{3}{\pi}\biggr)^{1/3}n_{\rm e}^{4/3}.
\end{eqnarray}
The Fermi momentum $p_{\rm F}$ is related to the electron number density by
$p_{\rm F}$\,$=$\,$\hbar(3\pi^2 n_{\rm e})^{1/3}$. $\theta_{\rm F}$ is defined
with the Fermi energy $\varepsilon_{\rm F}$ through the relation, $\varepsilon_{\rm F}
$\,$=$\,$\sqrt{m_{\rm e}^2c^4+p_{\rm F}^2c^2}=m_{\rm e}c^2\cosh\theta_{\rm F}$.
The Coulomb exchange term is evaluated with the Slater approximation,
while the direct term of the electrons' Coulomb energy vanishes because
of the charge neutrality condition.

\subsection{The Skyrme-Kohn-Sham equations}\label{Sec:KS}

From an appropriate functional derivative, one can derive the
Skyrme-Kohn-Sham equations for the single-particle wave functions:
\begin{equation}
\hat{h}^{(q)}(z)\psi_{\alpha\bs{k}}^{(q)}(\bs{r})
= \varepsilon_{\alpha\bs{k}}^{(q)}\psi_{\alpha\bs{k}}^{(q)}(\bs{r}),
\label{Eq:KS_psi}
\end{equation}
where the single-particle Hamiltonian is given by
\begin{eqnarray}
\hat{h}^{(q)}(z) &=&
-\bs{\nabla\cdot}\frac{\hbar^2}{2m_q^\oplus(z)}\bs{\nabla} + U^{(q)}(z) \nonumber\\[0.5mm]
&&+\; \frac{1}{2\zI}\bigl[ \bs{\nabla\cdot}\bs{I}^{(q)}(z)+\bs{I}^{(q)}(z)\bs{\cdot\nabla} \bigr].
\label{Eq:h_0}
\end{eqnarray}
Note that the differential operators in Eq.~(\ref{Eq:h_0}) act all spatial functions
located on the right side of them. $\varepsilon_{\alpha\bs{k}}^{(q)}$ represent
single-particle energies of the wave functions $\psi_{\alpha\bs{k}}^{(q)}$. For
each $\bs{k}$, the system exhibits discrete energy levels in a similar way as finite
nuclei, and each of those levels forms a ``band'' as a function of $\bs{k}$. We
refer to $\bs{k}$-dependent behavior of the energy levels as band structure.

The ``microscopic'' effective mass, $m_q^\oplus(z)$ (which should be distinguished
from the ``macroscopic'' one which will be discussed in Sec.~\ref{Sec:comparison}),
the time-even and time-odd mean-field potentials, $U^{(q)}(z)$ and $\bs{I}^{(q)}(z)$,
are defined, respectively, as follows:
\begin{eqnarray}
\frac{\hbar^2}{2m_q^\oplus(z)}
&=& \frac{\hbar^2}{2m_q} + \sum_{q'={\rm n,p}}C_{q'}^{\tau(q)}n_{q'}(z),
\label{Eq:h2M*}\\[1.5mm]
U^{(q)}(z) &=& \sum_{q'={\rm n,p}} \Bigl[
2C_{q'}^{\rho(q)}n_{q'}(z) + 2C_{q'}^{\triangle\rho(q)} \partial_z^2 n_{q'}(z)
\label{Eq:U_q}\nonumber\\[-2mm]
&&\hspace{8.3mm}+\;
C_{q'}^{\tau(q)}\tau_{q'}(z) + 2n_0^\gamma(z) C_{q'{\rm D}}^{\rho(q)}n_{q'}(z)
\Bigr]
\nonumber\\[0.5mm]
&&\hspace{8.3mm}+\;
\gamma n_0^{\gamma-1}(z)\sum_{t=0,1}C_{t{\rm D}}^\rho n_t^2(z) \nonumber\\
&&\hspace{8.3mm}+\;
U_{\rm Coul}(z)\delta_{q{\rm p}},
\\[2.5mm]
\bs{I}^{(q)}(z)
&=& -2 \sum_{q'={\rm n,p}}C_{q'}^{\tau(q)}\bs{j}_{q'}(z),
\end{eqnarray}
where
\begin{equation}
U_{\rm Coul}(z) = V_{\rm Coul}(z) -e^2\biggl(\frac{3}{\pi}\biggr)^{1/3}n_{\rm p}^{1/3}(z).
\end{equation}
For the sake of compactness, we have introduced a shorthand notation,
\begin{eqnarray}
C_{q'}^{{\rm X}(q)} &\equiv& \left\{
\begin{array}{cc}
C_0^{\rm X}+C_1^{\rm X} & (q=q'),\\[2mm]
C_0^{\rm X}-C_1^{\rm X} & (q\neq q'),
\end{array}\right.
\end{eqnarray}
where X stands for the superscript of the coefficients, i.e., $\rho$, $\tau$, or
$\Delta\rho$. Note that the time-odd potential is zero in a static situation.

In practice, one can work with the Skyrme-Kohn-Sham equations for
the Bloch orbitals $u_{\alpha\bs{k}}^{(q)}(z)$. Namely, by inserting
the explicit expression of Eq.~(\ref{Eq:spwf}) into Eq.~(\ref{Eq:KS_psi}),
we obtain:
\begin{equation}
\Bigl( \hat{h}^{(q)}(z) + \hat{h}_{\bs{k}}^{(q)}(z) \Bigr)u_{\alpha\bs{k}}^{(q)}(z)
= \varepsilon_{\alpha\bs{k}}^{(q)}u_{\alpha\bs{k}}^{(q)}(z),
\label{Eq:KS_u}
\end{equation}
where
\begin{equation}
\hat{h}_{\bs{k}}^{(q)}(z) \equiv \frac{\hbar^2\bs{k}^2}{2m_q^\oplus(z)} + \hbar\bs{k\cdot}\hat{\bs{v}}^{(q)}(z).
\label{Eq:h_k}
\end{equation}
Here, $\hat{\bs{v}}^{(q)}(z)$ is the so-called velocity operator defined by
\begin{eqnarray}
\hat{\bs{v}}^{(q)}(z)
&\equiv& \frac{1}{\zI\hbar}\bigl[ \bs{r},\hat{h}^{(q)}(z) \bigr] \nonumber\\[1mm]
&=& -\zI\hbar\Bigl(\bs{\nabla}\frac{1}{2m_q^\oplus(z)}+\frac{1}{2m_q^\oplus(z)}\bs{\nabla}\Bigr)
+ \frac{1}{\hbar}\bs{I}^{(q)}(z).
\label{Eq:velocity_op}
\hspace{7mm}
\end{eqnarray}
The terms that contain $\hat{v}_x^{(q)}$ and $\hat{v}_y^{(q)}$ do not contribute in
the present case. The additional $\bs{k}$-dependent term (\ref{Eq:h_k}) arises due to
differential operations on the factor $e^{\zI\bs{k\cdot r}}$. The commutator in the velocity
operator extracts the additional terms originated from the differential operation on $e^{\zI\bs{k\cdot r}}$,
since terms that commute with the coordinate $\bs{r}$ vanish. In this way, while we deal
with a 3D infinite system of crystalline slabs, the equations to be solved becomes essentially
one-dimensional ones, for the Bloch orbitals along $z$ direction, Eq.~(\ref{Eq:KS_u}).

We note that in the recent work of Ref.~\cite{Kashiwaba(2019)}, a functional
of the Barcelona-Catania-Paris-Madrid (BCPM) family \cite{Baldo(2013)} was
employed. Since the BCPM functional contains the kinetic density $\tau_q(\bs{r})$
only in the kinetic term, the resulting Kohn-Sham equations were greatly simplified.
In addition, since the microscopic effective mass is equal to the bare nucleon mass in
the BCPM functional used in Ref.~\cite{Kashiwaba(2019)}, the Bloch orbitals were
independent from $k_\parallel$, that substantially reduced the number of single-particle
orbitals. (It is to mention that there exists a variant of the BCPM functional with the
microscopic effective mass \cite{Baldo(2017)}.) The formulas given above are more
general ones, which can be readily extended for more complex crystals with higher
dimensions and are based on a widely-used Skyrme-type EDF which utilizes the
density-dependent microscopic effective mass, $m_q^\oplus$.

\subsection{Real-space TDDFT for solids in the velocity gauge}\label{Sec:TDDFT}

As described in Sec.~\ref{Sec:Mslab}, we will extract the collective mass of a slab
from a dynamic response of the system under a uniform external electric field,
$\bs{E}$\,$=$\,$E_z\hat{\bs{e}}_z$. The situation resembles the one encountered
in solid-state physics. Since a typical wavelength of an oscillatory electric field of light is,
in general, much longer than the size of an atomic unit cell, the electric field is often treated
as constant over the unit cell, known as the dipole approximation. One may express a
constant electric field by an external scalar potential, $\phi_{\rm ext}(\bs{r})$\,$=$\,$-E_z z$.
This expression is known as the length gauge, where a linear scalar potential is entirely
responsible for the electric field which couples with the coordinate by $-\bs{E\cdot r}$.
Time evolution of the system may be described by the time-dependent Kohn-Sham (TDKS)
equations for the Bloch orbitals in the length gauge,
\begin{equation}
\zI\hbar\frac{\partial u_{\alpha\bs{k}}^{(q)}(z,t)}{\partial t}
= \Bigl( \hat{h}^{(q)}(z,t) + \hat{h}_{\bs{k}}^{(q)}(z,t) -eE_z z \Bigr)u_{\alpha\bs{k}}^{(q)}(z,t).
\label{TDKS-l}
\end{equation}
The single-particle Hamiltonians have exactly the same form as described in
Sec.~\ref{Sec:KS} (except the time dependence of the densities and the potentials).
However, this external potential clearly violates the periodicity assumption and, hence,
cannot be used together with the Bloch boundary condition.

We can avoid this problem by employing a gauge transformation for the Bloch orbitals
to the so-called velocity gauge. Following the method developed by Bertsch et
al.~\cite{Bertsch(2000),Yabana(2012)}, we introduce a gauge transformation
for the Bloch orbitals from the length gauge to the velocity gauge,
\begin{equation}
\widetilde{u}_{\alpha\bs{k}}^{(q)}(z,t)
= \exp\biggl[-\frac{\zI e}{\hbar c}A_z(t)z\biggr]\,u_{\alpha\bs{k}}^{(q)}(z,t),
\label{Eq:u_tilde}
\end{equation}
and correspondingly for the single-particle Hamiltonian,
\begin{eqnarray}
\widetilde{h}^{(q)}(z,t)
= \exp\biggl[-\frac{\zI e}{\hbar c}A_z(t)z\biggr] \hat{h}^{(q)}(z,t) \exp\biggl[\frac{\zI e}{\hbar c}A_z(t)z\biggr].
\nonumber\\[0.5mm]\label{Eq:spH-v}
\end{eqnarray}
Here, $A_z(t)$ represents a time-dependent, spatially uniform vector potential which
is equivalent to the uniform external electric field by $E_z(t)$\,$=$\,$-(1/c)\,{\rm d}A_z(t)/{\rm d}t$,
and $c$ denotes the speed of light. The transformation may also be regarded as a Galilean
boost for the Bloch orbitals, $e^{\zI\Delta p_z z/\hbar}$, with $\Delta p_z$\,$=$\,$\int_0^t eE_z(t')\,{\rm d}t'$,
which is consistent with the classical picture of the acceleration associated with the external force.
Since the vector potential is constant in space, it does not violate the periodicity assumption.
(see Appendix~\ref{App:gauges} for details of the gauge transformations.)

It is easy to understand that the gauge transformation of the single-particle Hamiltonian
\eqref{Eq:spH-v} affects only the $\bs{k}$-dependent part, Eq.~\eqref{Eq:h_k}.
Namely, the TDKS equations in the velocity gauge take the following form:
\begin{equation}
\zI\hbar\frac{\partial\widetilde{u}_{\alpha\bs{k}}^{(q)}(z,t)}{\partial t}
= \biggl( \hat{h}^{(q)}(z,t) + \hat{h}_{\bs{k}(t)}^{(q)}(z,t) \biggr) \widetilde{u}_{\alpha\bs{k}}^{(q)}(z,t).
\label{Eq:TDKS-v}
\end{equation}
Notice that the Bloch wave vector in the single-particle Hamiltonian in
Eq.~(\ref{Eq:TDKS-v}) is now shifted as a function of time, according to:
\begin{equation}
\bs{k}(t) = \bs{k} + \frac{e}{\hbar c}A_z(t)\hat{\bs{e}}_z.
\label{Eq:k(t)}
\end{equation}
All densities can be expressed in terms of the Bloch orbitals in the velocity gauge,
$\widetilde{u}_{\alpha\bs{k}}^{(q)}(\bs{r},t)$, by inserting its inverse transform,
$u_{\alpha\bs{k}}^{(q)}(\bs{r},t)=\exp[\zI eA_z(t)z/\hbar c]\,\widetilde{u}
_{\alpha\bs{k}}^{(q)}(\bs{r},t)$, into Eqs.~\eqref{Eq:n_q}--\eqref{Eq:j_q}.
Thus, it is possible to work with the same form of Eqs.~\eqref{Eq:n_q}--\eqref{Eq:j_q},
simply replacing $u_{\alpha\bs{k}}^{(q)}(\bs{r},t)\rightarrow\widetilde{u}_{\alpha\bs{k}}^{(q)}(\bs{r},t)$
with the time dependent shift for the Bloch wave number $k_z\rightarrow k_z(t)$
according to Eq.~\eqref{Eq:k(t)} in the kinetic and the momentum densities.
It is to mention that additional care should be taken for the single-particle
Hamiltonian in the velocity gauge, when one works with a non-local potential
(see, e.g., Ref.~\cite{Yabana(2006)}).

\subsection{Real-time method for the collective mass}\label{Sec:Mslab}

In the inner crust of a neutron star, where nuclear bundles and a sea of dripped
neutrons coexist, it is not obvious how to distinguish bound and unbound (free)
neutrons. If a naive picture is adopted, one may distinguish them according
either to magnitude of the single-particle energies and the nuclear potential,
or to density distributions \cite{Papakonstantinou(2013)}. However, we do
not expect such a clear separation in reality. Due to the Bragg scattering,
part of ``free'' neutrons may be entrained (or, say, ``effectvely bound'')
to the periodic potential \cite{Chamel(2017)}. Here, we introduce a real-time
method to extract the collective mass of a nuclear bundle (not necessarily a slab)
immersed in a sea of dripped neutrons, which directly measures the number
of the effectively bound (bound\,+\,entrained) nucleons.

\begin{figure} [t]
\includegraphics[width=7.8cm]{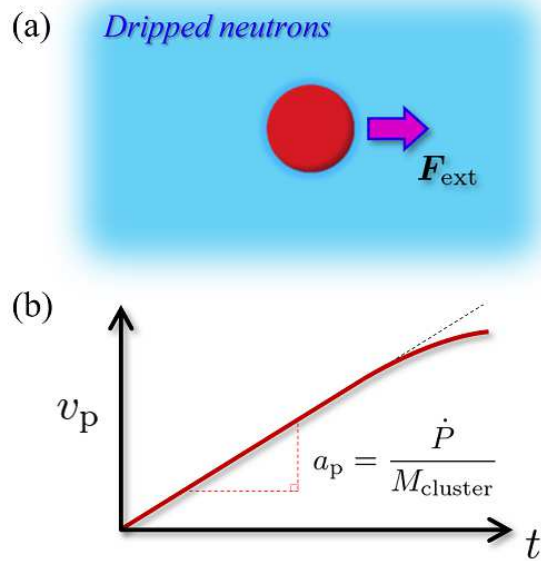}\vspace{-2mm}
\caption{
Schematic picture showing the idea of the real-time method for extracting
the collective mass of a nuclear cluster immersed in a sea of dripped neutrons.
(a):~We consider a situation where a nuclear cluster is embedded in background
neutron matter. We exerts external force $F_{\rm ext}$ on localized protons
inside the cluster. (b):~Expected time evolution of the velocity of the center-of-mass
position of protons $v_{\rm p}(t)$, showing constant acceleration motion with
$a_{\rm p}$\,$=$\,$\dot{P}/M_{\rm cluster}$. In case where motion is too
fast and some excitation modes are induced, we may observe deceleration as
indicated in the right-top part of the $t$-$v_{\rm p}$ plot.
}\vspace{-3mm}
\label{Fig:idea}
\end{figure}

We employ a simple idea as depicted in Fig.~\ref{Fig:idea}. We exerts external force
$F_{\rm ext}$ on localized protons inside a nuclear cluster (a slab, in the present case),
see Fig.~\ref{Fig:idea}(a). The cluster then moves with constant acceleration $a_{\rm p}$
according to the classical relation, $\dot{P}$\,$=$\,$M_{\rm cluster}\,a_{\rm p}$,
where $\dot{P}$ is the rate of change of the linear momentum and $M_{\rm cluster}$
is the collective mass of the cluster. Here and henceforth, a dot on a quantity represents
its time derivative. Thus, by measuring the acceleration $a_{\rm p}$ and the rate of
change of the linear momentum $\dot{P}$, one can deduce the collective mass simply
by $M_{\rm cluster}$\,$=$\,$\dot{P}/a_{\rm p}$, see Fig.~\ref{Fig:idea}(b). We note
that the force should be small enough so as to avoid excitations of the system that would
result in deviation from the uniform acceleration motion, as indicated in the right-top
part of Fig.~\ref{Fig:idea}(b). It is to mention here that if superfluidity is present, which
is neglected in the present work, it would prevent the system from exciting until the motion exceeds
certain critical velocity for breaking Cooper pairs, whereas additional collective excitation modes,
the Anderson-Bogoliubov phonons, may be present \cite{Inakura(2017),Inakura(2019)}.

\begin{table*}[t]
\caption{
List of quantities that characterize static properties of the systems. In the list,
$Y_{\rm p}$ is the proton fraction, $a$ is the slab period in fm, $\mu_q$ is the Fermi
energy in MeV, $U_0^{(q)}$ is the maximum value of the nuclear potential in MeV,
$\bar{n}_q$ is the average nucleon number density in fm$^{-3}$, $n_{\rm n}
^{\rm bg}$ is the background neutron number density in fm$^{-3}$, which is the
density at the midpoint between the slabs, i.e. $n_{\rm n}^{\rm bg}=n_{\rm n}(0)=n_{\rm n}(a)$
in the present case, $n_{\rm n}^{\rm f}$ [$=\bar{n}_{\rm n}-n_{\rm n}^{\rm e.b.}$,
where $n_{\rm n}^{\rm e.b.}$\,$=$\,$N_{\rm n}^{\rm e.b.}/a$; cf.\ Eq.~(\ref{Eq:def_N_local})]
is the number density of ``free'' (or ``energetically unbound'') neutrons, and $m_{q,{\rm bg}}
^\oplus$\,$=$\,$m_q^\oplus[n_{\rm bg}]$ is the microscopic effective mass at the
background neutron number density $n_{\rm bg}^{\rm n}$. Quantities with an index
$q$ are those for neutrons ($q$\,$=$\,${\rm n}$) or protons ($q$\,$=$\,${\rm p}$).
For the bare neutron and proton masses, $m_{\rm n}$\,$=$\,939.565~MeV/$c^2$
and $m_{\rm p}$\,$=$\,938.272~MeV/$c^2$ were used, respectively.
}\vspace{-1mm}
\label{Table:summary}
\begin{center}
\begin{tabular*}{\textwidth}{@{\extracolsep{\fill}}cccccccccccc}
\hline\hline
$Y_{\rm p}$ & $a$ (fm) & $\mu_{\rm n}$ (MeV) & $\mu_{\rm p}$ (MeV) & $U_0^{({\rm n})}$ (MeV) & $U_0^{(\rm p)}$ (MeV) & $\bar{n}_{\rm n}$ (fm$^{-3}$) & $\bar{n}_{\rm p}$ (fm$^{-3}$) & $n_{\rm n}^{\rm bg}$ (fm$^{-3}$) & $n_{\rm n}^{\rm f}$ (fm$^{-3}$) & $m_{\rm n,bg}^\oplus/m_{\rm n}$ & $m_{\rm p,bg}^\oplus/m_{\rm p}$ \\
\hline
0.5 & 20 & $-$16.8 & $-$12.1 & $-$2.10\,$\times$\,10$^{-3}$ & $-$5.32 & 0.020 & 0.020 & 4.41\,$\times$\,10$^{-8}$ & 0 & 1.000 & 1.000 \\
0.4 & 22 & $-$7.11 & $-$22.1 & $-$3.93\,$\times$\,10$^{-5}$ & $-$5.26 & 0.024 & 0.016 & 3.29\,$\times$\,10$^{-8}$ & 0 & 1.000 & 1.000 \\
0.3 & 25 & 0.15 & $-$34.1 & $-$2.49\,$\times$\,10$^{-2}$ & $-$5.36 & 0.028 & 0.012 & 8.18\,$\times$\,10$^{-5}$ & 5.86\,$\times$\,10$^{-6}$ & 1.000 & 1.000 \\
0.2 & 26 & 4.80 & $-$43.5 & $-$2.69 & $-$16.8 & 0.032 & 0.008 & 7.04\,$\times$\,10$^{-3}$ & 4.06\,$\times$\,10$^{-3}$& 0.973 & 0.989 \\
0.1 & 33 & 7.83 & $-$51.7 & $-$8.25 & $-$34.3 & 0.036 & 0.004 & 2.05\,$\times$\,10$^{-2}$ & 1.30\,$\times$\,10$^{-2}$ & 0.925 & 0.969 \\
\hline\hline
\end{tabular*}\vspace{-3mm}
\end{center}
\end{table*}

We apply the real-time method for the slab phase of nuclear matter. With the method
described in Sec.~\ref{Sec:TDDFT}, the above mentioned situation can be realized
numerically with a virtual external electric field, $\bs{E}$\,$=$\,$E_z\hat{\bs{e}}_z$, that
couples only with protons inside a slab. From time evolution, the position of the slab
can be traced through the expectation value of the center-of-mass position of protons,
$Z_{\rm p}(t)$\,$=$\,$\frac{1}{N_{\rm p}}\int_0^a z n_{\rm p}(z,t)\,{\rm d}z$,
and the acceleration is given by $a_{\rm p}(t)$\,$=$\,${\rm d}^2Z_{\rm p}(t)/{\rm d}t^2$.
We then define the collective mass (per unit area) of the slab, $M_{\rm slab}$,
as well as that of protons, $M_{\rm p}$, respectively, as follows:
\begin{eqnarray}
M_{\rm slab}(t) &=& \frac{\dot{P}_{\rm tot}(t)}{a_{\rm p}(t)},\label{Eq:M_slab}\\[0.5mm]
M_{\rm p}(t) &=& \frac{\dot{P}_{\rm p}(t)}{a_{\rm p}(t)},\label{Eq:M_p}
\end{eqnarray}
where the linear momentum (per unit area) can be computed from
the momentum densities as
\begin{equation}
P_q(t) = \hbar\int_0^a j_{z,q}(z,t)\,{\rm d}z.
\end{equation}
Here, $P_{\rm tot}$ denotes the sum of the neutron and proton linear
momenta, $P_{\rm tot}$\,$=$\,$P_{\rm n}+P_{\rm p}$. The rate
of change of the total momentum coincides with the total force exerting
on the system, $\dot{P}_{\rm tot}(t)$\,$=$\,$eN_{\rm p}E_z(t)$.

In this way, in principle, one can deduce the collective mass of a nuclear cluster
immersed in a sea of dripped neutrons from a time-dependent response of the
system to the external force. From the collective mass of a cluster, one can
estimate the number of effectively bound (bound\,+\,entrained) neutrons,
which, in turn, defines the rest---the conduction neutrons---moving freely in
the crystalline potential. We will discuss this point in Sec.~\ref{Sec:comparison}.

\subsection{Computational details}\label{Sec:details}

The Bloch orbitals along $z$ coordinate (normal to the slab) are represented
on uniform grid points in real space. The grid spacing is set to $\Delta z$\,$=$\,0.5\,fm\footnote{
\add{
To decide the grid spacing $\Delta z$, we carried out static calculations for a neutron-dripped
system (the $n_\mathrm{B}$\,=\,0.4, $Y_\mathrm{p}$\,=\,0.1 case), changing $\Delta z$\,$\simeq$\,0.2--1.0\,fm.
We confirmed that the total energy per nucleon ($E/A$) is converging with decreasing $\Delta z$
and found that $\Delta z$\,=\,0.5\,fm already achieves a 0.01\% (0.1\,keV) accuracy.
Based on this observation, we decided to set $\Delta z$\,=\,0.5\,fm in the present work.
}
}.
First and second spatial derivatives are evaluated with the 9-point finite difference
formulas. The Coulomb potential is computed with fast Fourier transforms. We discretize
$k_\parallel$ below the Fermi energy with a $\Delta k_\parallel=0.01$-fm$^{-1}$ step.
In most cases, however, the Fermi energy lies in between the discretized values, i.e.
$\varepsilon_{\alpha k_z k_\parallel^j}^{(q)}$\,$<$\,$\mu_q$\,$<$\,$\varepsilon_{\alpha k_z k_\parallel^{j+1}}^{(q)}$
($k_\parallel^j \equiv j\Delta k_\parallel$). We thus add $\delta k_\parallel=k_\parallel^{\rm max}-j\Delta k_\parallel$
to evaluate the integral over $k_\parallel$ so that $\varepsilon_{\alpha k_z k_\parallel^{\rm max}}^{({\rm n})}=\mu_{\rm n}$
is ensured. The first Brillouin zone, $-\pi/a$\,$\le$\,$k_z$\,$\le$\,$\pi/a$, is discretized into
$N_{k_z}$ points. Note that actual calculations were carried out for $-\pi/a,\dots,0\dots,\pi/a$,
i.e. $N_{k_z}+1$ points, where $k_z$\,$=$\,$\pm\pi/a$ are treated equally with a factor 0.5,
which is necessary to ensure $\bs{j}_q$\,$=$\,$0$ [cf.\ Eq.~\eqref{Eq:j_q}] for static solutions.
For the static calculations, we found that $N_{k_z}$\,$\ge$\,10 provides results with negligible
spurious finite-size effects. We note, however, that for dynamic calculations $N_{k_z}$\,$\ge$\,40
turned out to be necessary to obtain convergent results, especially for neutron dripped systems.
We thus set $N_{k_z}$\,$=$\,40 in the present paper. A static solution of the Kohn-Sham
equations is obtained by iterative diagonalizations of the single-particle Hamiltonian
matrix for each set of $\bs{k}$\,$=$\,$(k_\parallel,k_z)$ until convergence
of $\Delta E$\,$<$\,10$^{-14}$\,MeV (energy change from the previous iteration)
is achieved. For the time integration, we employ the 6th-order Taylor expansion
method with a single predictor-corrector step. The time step is set to
$\Delta t$\,$=$\,0.1\,fm/$c$. In the calculations presented in this paper,
the Coulomb exchange term in Eq.~(\ref{Eq:E_Coul_proton}) has been
neglected, which is expected to be a minor correction.

In practice, static calculations are performed without the external potential. We then
dynamically switch on the uniform external electric field smoothly as a function of time,
i.e. $E_z(t)$\,$=$\,$s(t,T,\eta)E_z$ for 0\,$\le$\,$t$\,$\le$\,$T$, where $s(t,T,\eta)$
is a switching function which varies smoothly from 0 to 1:
\begin{equation}
s(t,T,\eta) = \frac{1}{2} + \frac{1}{2}\tanh\Bigl[ \eta\tan\Bigl(\frac{\pi t}{T}-\frac{\pi}{2}\Bigr) \Bigr],
\end{equation}
where $\eta$ and $T$ are parameters that control the speed and the duration of the
variation, respectively. In the present study, $\eta$\,$=$\,2 and $T$\,$=$\,1,500\,fm/$c$
are used for isolated slabs (the $Y_{\rm p}$\,$=$\,0.5 and 0.4 cases), while $\eta$\,$=$\,1
and $T$\,$=$\,2,000\,fm/$c$ are used for neutron dripped systems (the $Y_{\rm p}$\,$=$\,0.3,
0.2, and 0.1 cases). Without the switching function, we found that the electric field induces
an isovector-type oscillation mode (with a period of $\approx 200$\,fm/$c$). The strength
of the external force (per unit area) is set to $F_z$\,$=$\,$eE_z$\,$=$\,$10^{-3}$\,MeV/fm$^3$.

\section{NUMERICAL RESULTS}\label{Sec:Results}

In this section, we present the results of static and dynamic self-consistent band
calculations for the slab phase of nuclear matter. Under a realistic $\beta$-equilibrium
condition, the slab phase may emerge at a bottom layer of the inner crust close to the
crust-core transition at baryon density, $\bar{n}_{\rm b}$\,$\simeq$\,0.07--0.09\,fm$^{-3}$,
with tiny proton fraction, $Y_{\rm p}$\,$\simeq$\,0.02--0.03 \cite{Kashiwaba(2019)}.
In this work, however, we restrict ourselves to the cases at a fixed baryon density,
$\bar{n}_{\rm b}$\,$=$\,0.04~fm$^{-3}$, with different proton fractions from
$Y_{\rm p}$\,$=$\,0.1 to 0.5. Those conditions allow us to investigate from isolated
slabs, where all nucleons are localized inside the slabs, to slabs immersed in a sea of
dripped neutrons, similar to the situation encountered in the inner crust of neutron stars.
Various values that characterize those systems are summarized in Table~\ref{Table:summary}.

\subsection{Isolated slabs: \textit{Y}$_\mathbf{p}$ = 0.5 and 0.4}\label{Sec:isolated}

\begin{figure} [t]
\includegraphics[width=8.2cm]{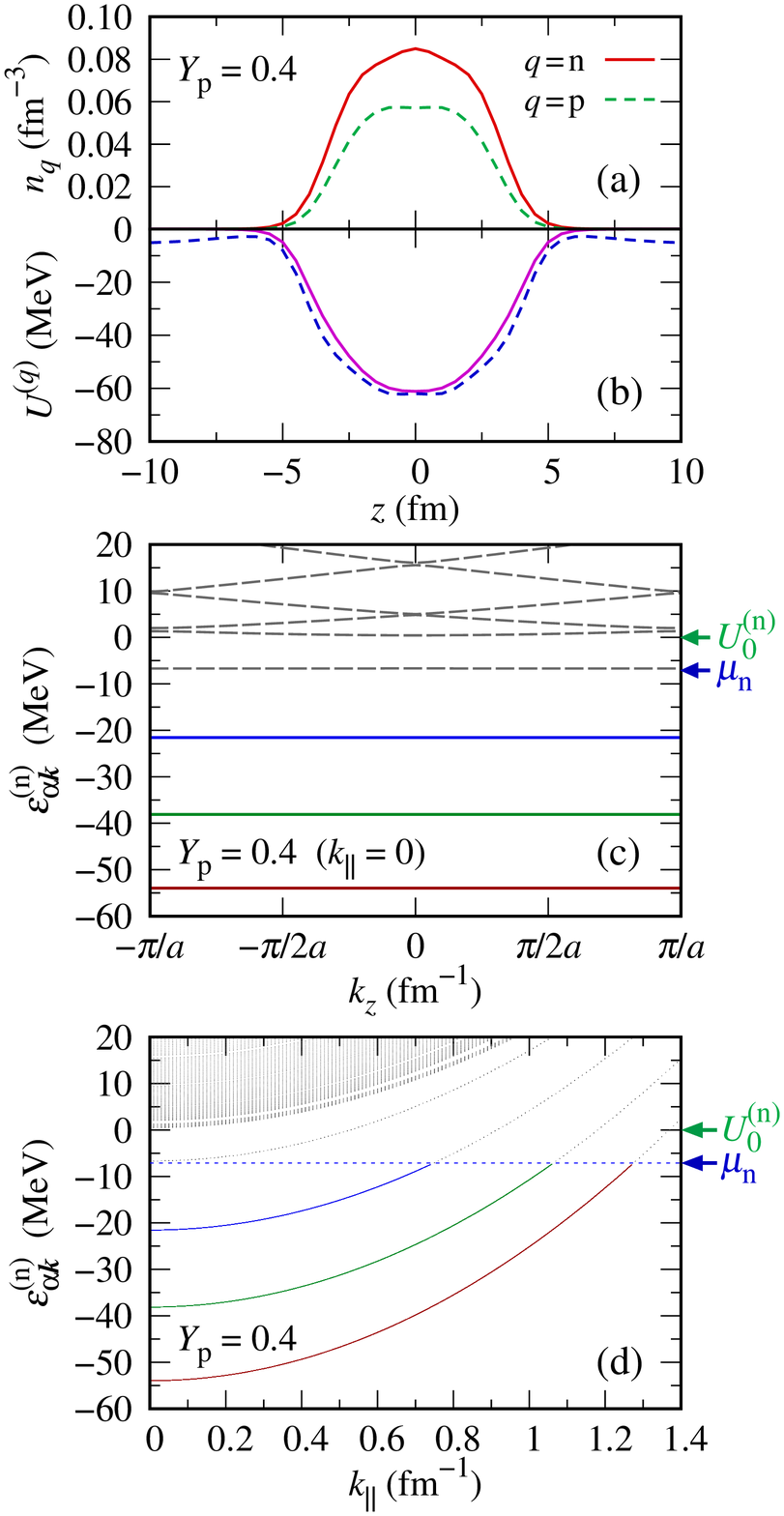}\vspace{-3mm}
\caption{
Results of static self-consistent band calculations for the system of isolated slabs with proton
fraction $Y_{\rm p}$\,$=$\,0.4 at baryon density $\bar{n}_{\rm b}$\,$=$\,0.04\,fm$^{-3}$.
In panels (a) and (b), density distributions and the mean-field potentials are shown, respectively,
as a function of $z$ coordinate, for neutrons (solid line) and for protons (dashed lines). In panels
(c) and (d), neutron single-particle energies $\varepsilon_{\alpha\bs{k}}^{({\rm n})}$ are shown.
In panel (c), $\varepsilon_{\alpha\bs{k}}^{({\rm n})}$ for $k_\parallel=0$ are shown as a function
of the Bloch wave number $k_z$ within the first Brillouin zone, $-\pi/a$\,$\le$\,$k_z$\,$\le$\,$\pi/a$.
In panel (d), $\varepsilon_{\alpha\bs{k}}^{({\rm n})}$ are shown as a function of $k_\parallel$
(energies for all $k_z$ values are shown). In those panels, the neutron Fermi energy $\mu_{\rm n}$
and the maximum value of the mean-field potential for neutrons $U_0^{({\rm n})}$ are indicated by
arrows. Occupied levels are shown by solid lines, while unoccupied levels are also indicated by dashed lines.
}\vspace{-4mm}
\label{Fig:static_Yp04}
\end{figure}

\begin{figure} [t]
\includegraphics[width=7.8cm]{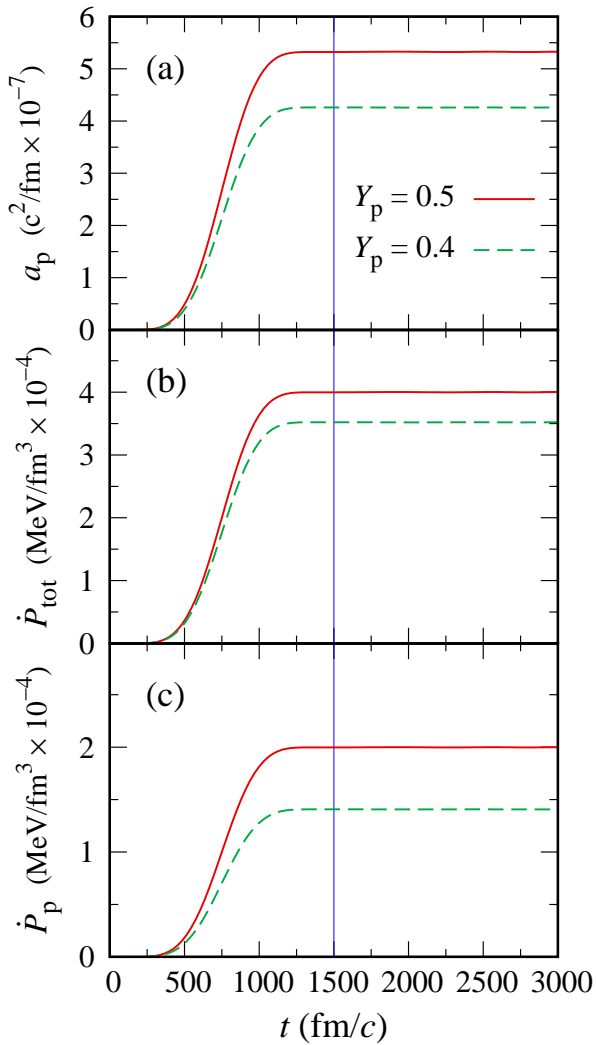}\vspace{-1mm}
\caption{
Results of TDDFT calculations for the systems of isolated slabs with proton fractions $Y_{\rm p}$\,$=$\,0.5
(solid line) and 0.4 (dashed line) at baryon density $\bar{n}_{\rm b}$\,$=$\,0.04\,fm$^{-3}$.
In panel (a), acceleration of the center-of-mass position of protons, $a_{\rm p}(t)$, is shown, while
time derivative of the linear momentum (per unit area) of the whole system, $\dot{P}_{\rm tot}(t)$,
and that of protons, $\dot{P}_{\rm p}(t)$, are shown in panels (b) and (c), respectively, as a function
of time. The vertical line indicates the time up to which the external force is turned on.
}
\label{Fig:a_p_Yp04-05}
\end{figure}

\begin{figure} [t]
\includegraphics[width=8cm]{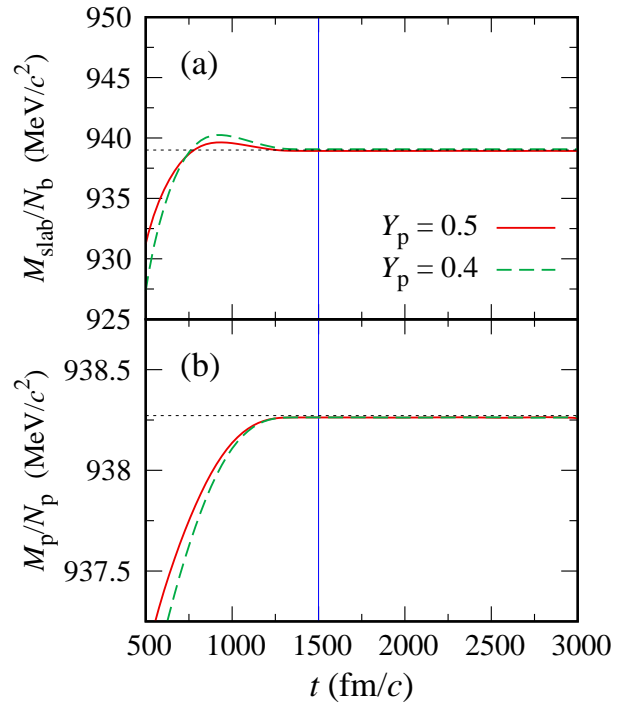}\vspace{-1mm}
\caption{
Results of TDDFT calculations for the systems of isolated slabs with proton fractions $Y_{\rm p}$\,$=$\,0.5
(solid line) and 0.4 (dashed line) at baryon density $\bar{n}_{\rm b}$\,$=$\,0.04\,fm$^{-3}$.
In panels (a) and (b), the collective mass per nucleon of a slab, $M_{\rm slab}(t)/N_{\rm b}$,
and that of protons, $M_{\rm p}(t)/N_{\rm p}$, are shown, respectively, as a function of time.
The vertical line indicates the time up to which the external force is turned on. The horizontal
dotted line in panels (a) and (b) represents the average bare nucleon mass and the bare proton
mass, respectively.
}\vspace{-3mm}
\label{Fig:Mcoll_Yp04-05}
\end{figure}

As a proof of principle, let us first investigate isolated slabs in a vacuum, i.e.\ without
dripped neutrons, that correspond to the $Y_{\rm p}$\,$=$\,0.5 and 0.4 cases. To find the
optimal slab period $a$, static band calculations were carried out for various values of $a$.
To the accuracy determined by the grid spacing, we find that $a$\,$=$\,20\,fm and 22\,fm
correspond to the lowest energy configuration for $Y_{\rm p}$\,$=$\,0.5 and 0.4, respectively.
As an example, we show in Fig.~\ref{Fig:static_Yp04} some static properties of the system
with $Y_{\rm p}$\,$=$\,0.4. Figure~\ref{Fig:static_Yp04}(a) shows density profiles
of neutrons (solid line) and protons (dashed line) as a function of $z$ coordinate normal
to the slab. In this case, both neutrons and protons are localized within the slab. The
density distributions exhibit `neutron skin,' analogous to a neutron-rich finite nucleus.
In Fig.~\ref{Fig:static_Yp04}(b), the mean-field potentials, $U^{(q)}(z)$, for neutrons (solid
line) and protons (dashed line) are shown as a function of $z$ coordinate. While the global
structure is similar to one another, the potential for protons is slightly deeper than that for neutrons,
because of the neutron richness and the presence of background electrons. Since occupied orbitals
in this case are well localized, the Bragg scattering does not play any role and the system does
not manifest band structure. It is visible in Fig.~\ref{Fig:static_Yp04}(c), where neutron
single-particle energies are shown as a function of the Bloch wave number $k_z$ within
the first Brillouin zone, $-\pi/a$\,$\le$\,$k_z$\,$\le$\,$\pi/a$, for $k_\parallel=0$.
In the figure, occupied levels are represented by solid lines, while unoccupied levels are
indicated by dashed lines. The band index runs from the lowest level to higher ones, i.e.
$\varepsilon_{1,k_z}^{({\rm n})}$\,$<$\,$\varepsilon_{2,k_z}^{({\rm n})}$\,$<$\,$\cdots$,
and the lowest three levels ($\alpha$\,$=$\,1, 2, and 3) are occupied in the present case.
The maximum value of the nuclear potential $U_0^{({\rm n})}$ and the neutron Fermi energy
$\mu_{\rm n}$ are indicated by arrows. Indeed, all the occupied levels in this case are within
the nuclear potential (i.e., $\mu_{\rm n}$\,$<$\,$U_0^{({\rm n})}$). As a result, those occupied
states are well localized, and there is essentially no $k_z$ dependence in the single-particle
energies. We note that unoccupied states with energies above $U_0^{({\rm n})}$, which
extend spatially outside the slabs, exhibit band structure. (We will come back to this point
when we analyze slabs with dripped neutrons in Sec.~\ref{Sec:dripped}.) Note that each
single-particle energy also has $k_\parallel$ dependence as shown in Fig.~\ref{Fig:static_Yp04}(d),
where $\varepsilon_{\alpha\bs{k}}^{({\rm n})}$ are shown as a function of $k_\parallel$.
Occupied states are those below the Fermi energy, indicted by a horizontal dashed line
(i.e. $\varepsilon_{\alpha\bs{k}}^{({\rm n})}<\mu_{\rm n}$).

Next, we show the results of time-dependent self-consistent band calculations for the $Y_{\rm p}
$\,$=$\,0.5 and 0.4 cases in Figs.~\ref{Fig:a_p_Yp04-05} and \ref{Fig:Mcoll_Yp04-05}. In both figures,
the results for the $Y_{\rm p}$\,$=$\,0.5 case are represented by solid lines, while those for the
$Y_{\rm p}$\,$=$\,0.4 case are represented by dashed lines. In Figs.~\ref{Fig:a_p_Yp04-05}(a),
\ref{Fig:a_p_Yp04-05}(b), and \ref{Fig:a_p_Yp04-05}(c), we show, respectively, acceleration
of protons, $a_{\rm p}(t)$, time derivative of the total linear momentum, $\dot{P}_{\rm tot}(t)$,
and time derivative of the proton linear momentum, $\dot{P}_{\rm p}(t)$, as a function of time.
As mentioned in Sec.~\ref{Sec:details}, we introduce the external force gradually as a function
of time for a time interval $t$\,$=$\,0--1500\,fm/$c$ (the end time is indicated by a vertical line).
It means that after $t$\,$=$\,1500\,fm/$c$, the force exerting on protons is constant. As can be
seen from the figure, the slabs in both systems show constant acceleration motion after
$t$\,$=$\,1500\,fm/$c$, as expected. By taking the ratios of those quantities, we can
obtain the collective masses, according to Eqs.~(\ref{Eq:M_slab}) and (\ref{Eq:M_p}).

Figure~\ref{Fig:Mcoll_Yp04-05} shows the extracted collective masses (per unit area) as a function
of time. In Figs.~\ref{Fig:Mcoll_Yp04-05}(a) and \ref{Fig:Mcoll_Yp04-05}(b), the collective mass
of the slab, $M_{\rm slab}(t)$, and that of protons, $M_{\rm p}(t)$, are presented, respectively.
To make a comparison of the results for different proton fractions, the collective masses are divided
by the total number of baryons $N_{\rm b}$ or protons $N_{\rm p}$ (per unit area) in the unit cell.
From the figure, we find that the extracted collective masses are fairly stable after $t$\,$=$\,1500\,fm/$c$,
where the external force is constant. The extracted masses agree very well with the average baryon
mass in Fig.~\ref{Fig:Mcoll_Yp04-05}(a) and with the bare proton mass in Fig.~\ref{Fig:Mcoll_Yp04-05}(b),
which are indicated by horizontal dotted lines, as they should be. We note that $M_{\rm slab}(t)$ and
$M_{\rm p}(t)$ show different behavior for $t$\,$<$\,1500\,fm/$c$, where the strength of the
external force is increasing as a function of time. It is because of the fact that the external force
exerts only on protons, and neutrons have to follow protons' motion dynamically through mutual
interactions that gives rise to a slight delay of its response to the force. The results clearly show
that our real-time method works properly for isolated slabs.

\subsection{Neutron-dripped slabs: \textit{Y}$_\mathbf{p}$ = 0.3, 0.2, and 0.1}\label{Sec:dripped}

In this section, we repeat the same analysis as in Sec.~\ref{Sec:isolated}
for the systems of slabs with dripped neutrons, i.e.\ the $Y_{\rm p}$\,$=
$\,0.3, 0.2, and 0.1 cases. From static band calculations, we find that the
optimal slab period to be $a$\,$=$\,25\,fm, 26\,fm, and 33\,fm for the
$Y_{\rm p}$\,$=$\,0.3, 0.2, and 0.1 cases, respectively. We note that
the $Y_{\rm p}$\,$=$\,0.3 case lies close to the border between isolated and
neutron-dripped systems. As the proton fraction decreases further, substantial
portion of neutrons are dripped out and the slabs are immersed in a sea of
dripped neutrons.

\begin{figure} [t]
\includegraphics[width=8.2cm]{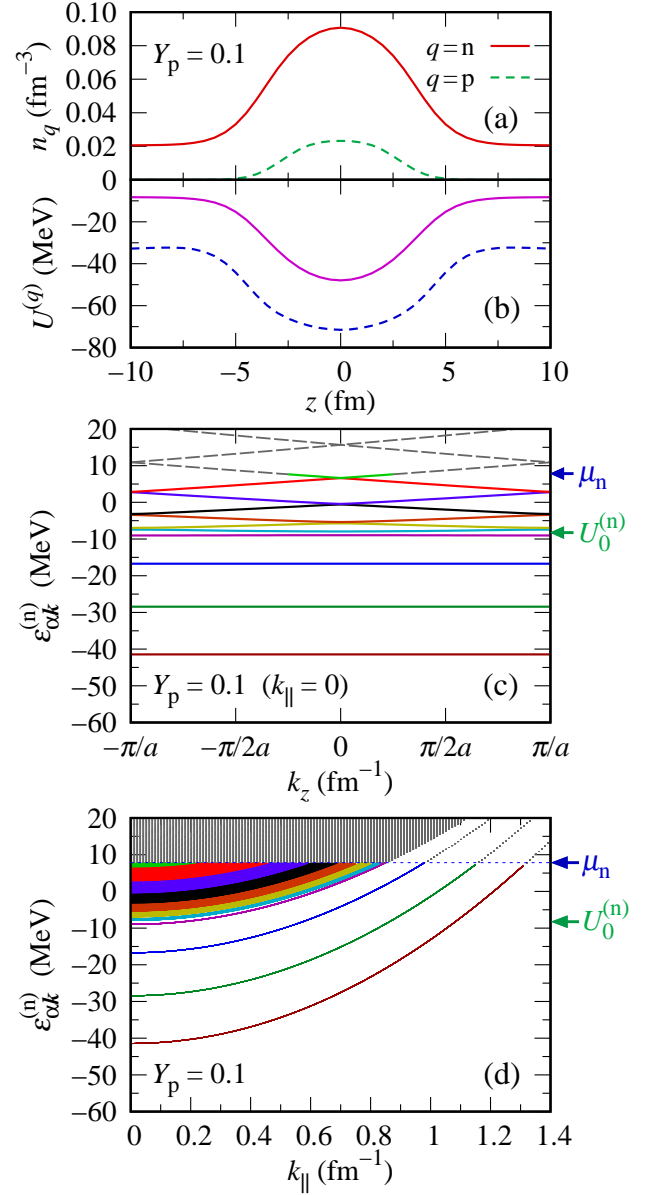}\vspace{-1mm}
\caption{
Same as Fig.~\ref{Fig:static_Yp04}, but for the system of slabs immersed in dripped neutrons
with proton fraction $Y_{\rm p}$\,$=$\,0.1 at baryon density $\bar{n}_{\rm b}$\,$=$\,0.04\,fm$^{-3}$.
In panels (c) and (d), solid lines show occupied levels, while unoccupied levels are also indicated by dashed lines.
}\vspace{-3mm}
\label{Fig:static_Yp01}
\end{figure}

As an illustrative example, we show in Fig.~\ref{Fig:static_Yp01}(a) density distributions of
neutrons (solid line) and protons (dashed line) as a function of $z$ coordinate for the $Y_{\rm p}$\,$=$\,0.1
case. Note that since the baryon density is fixed to $\bar{n}_{\rm b}$\,$=$\,0.04\,fm$^{-3}$,
the total number of protons decreases with the proton fraction. From the figure, it is evident that
there exist dripped neutrons outside the slab. Comparing with Fig.~\ref{Fig:static_Yp04}(a), the
width of the slab is slightly wider, which is partly due to the lesser number of protons that attract
neutrons to the slab. Since the dripped neutrons extend spatially over the whole space, they are
affected by the Bragg scattering off the periodic potential and the corresponding energy levels
manifest band structure.

\begin{figure} [t]
\includegraphics[width=7.8cm]{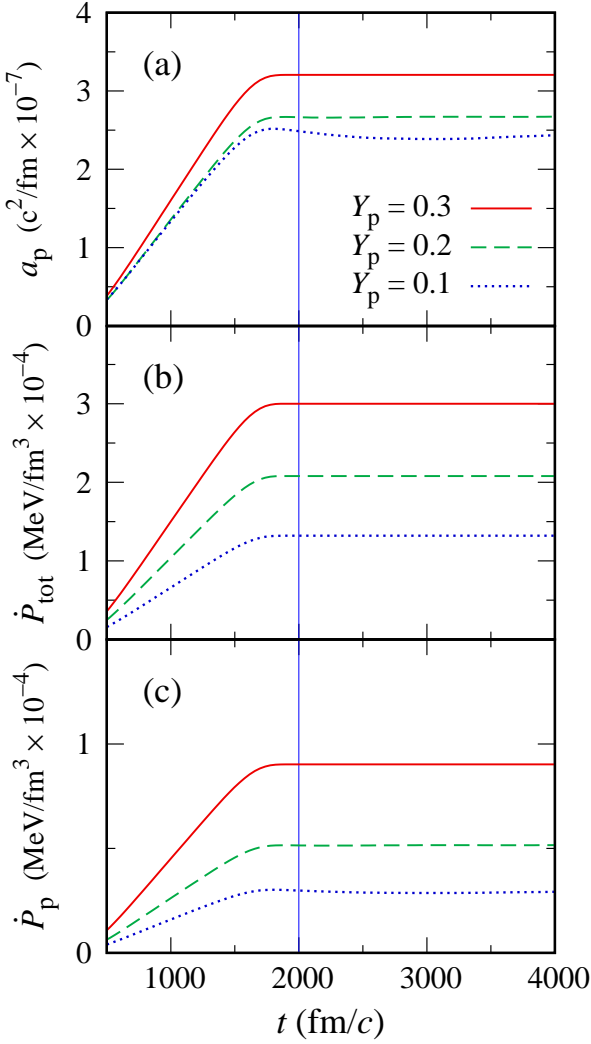}\vspace{-1mm}
\caption{
Same as Fig.~\ref{Fig:a_p_Yp04-05}, but for the systems of slabs immersed in dripped neutrons
with proton fractions $Y_{\rm p}$\,$=$\,0.3 (solid line), 0.2 (dashed line), and 0.1 (dotted line)
at baryon density $\bar{n}_{\rm b}$\,$=$\,0.04\,fm$^{-3}$. The vertical line indicates
the time up to which the external force is turned on.
}\vspace{-3mm}
\label{Fig:a_p_Yp01-03}
\end{figure}

\begin{figure} [t]
\includegraphics[width=8cm]{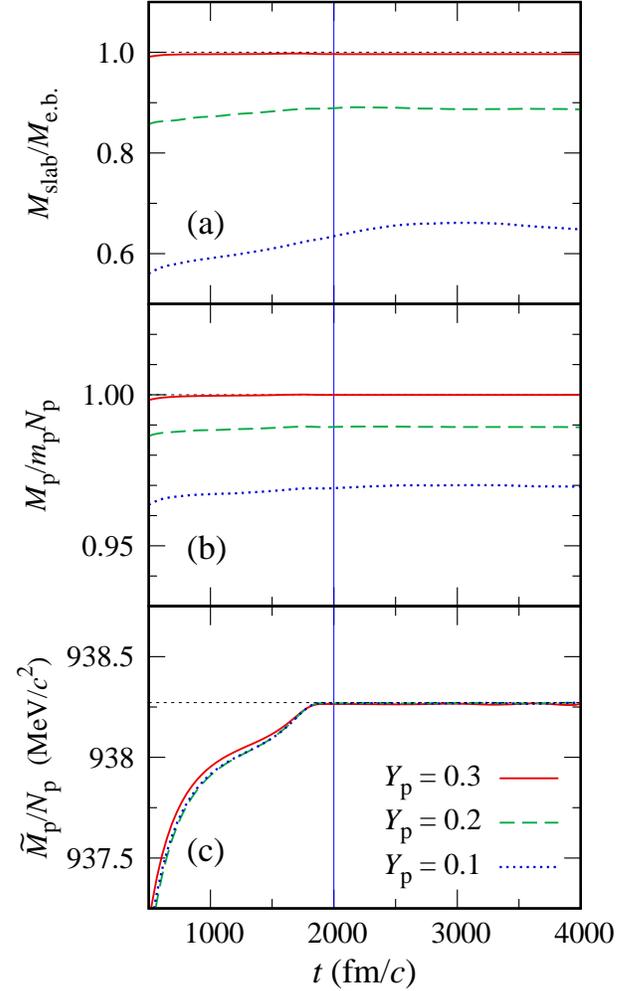}\vspace{-1mm}
\caption{
Results of TDDFT calculations for the systems of slabs immersed in dripped neutrons with proton
fractions $Y_{\rm p}$\,$=$\,0.3 (solid line), 0.2 (dashed line), and 0.1 (dotted line) at baryon
density $\bar{n}_{\rm b}$\,$=$\,0.04\,fm$^{-3}$ are shown as a function of time. In panel
(a), the ratio of the collective mass of a slab, $M_{\rm slab}(t)$, to the mass of ``energetically
bound'' nucleons, $M_{\rm e.b.}$, is shown. In panel (b), the ratio of the collective mass of
protons, $M_{\rm p}(t)$, to the total mass of protons, $m_{\rm p}N_{\rm p}$, is shown. In panel (c),
the collective mass of protons evaluated with a modified momentum density, $\widetilde{M}_{\rm p}(t)$\,$=$\,$
(1/a_{\rm p}(t)){\rm d}\widetilde{P}_{\rm p}(t)/{\rm d}t$, divided by the total number of protons,
is presented (See Sec.~\ref{Sec:M_p} for details). All the quantities are evaluated per unit area and it
does not affect their ratios. The vertical line indicates the time up to which the external force is turned on.
}\vspace{-3mm}
\label{Fig:Mcoll_Yp01-03}
\end{figure}

In Fig.~\ref{Fig:static_Yp01}(b), the mean-field potentials for neutrons (solid line) and
protons (dashed line) are shown as a function of $z$ coordinate for the $Y_{\rm p}$\,$=$\,0.1
case. The potential for neutrons (protons) is shallower (deeper) than the $Y_{\rm p}$\,$=$\,0.4
case shown in Fig.~\ref{Fig:static_Yp04}(b), due to smaller (lager) number of protons (neutrons).
A characteristic feature of a neutron-dripped system is that the nuclear potentials are negative even
outside the slabs. The potential for protons is much deeper than that for neutrons, while the
depth (i.e., difference between the minimum and the maximum values) is similar in magnitude
(roughly 40\,MeV in the present case).

In Fig.~\ref{Fig:static_Yp01}(c), we show the neutron single-particle energies as a function
of the Bloch wave number $k_z$ for $k_\parallel=0$. As in Fig.~\ref{Fig:static_Yp04}(c),
occupied levels are shown by solid lines, while unoccupied levels are indicated by dashed lines.
Note that the last occupied level ($\alpha$\,$=$\,11) is partially filled, which would allow neutrons
in that level to flow without a band gap even for the $z$ direction. The states below the maximum
value of the nuclear potential $U_0^{({\rm n})}$ are localized inside the slabs and their energies
have no $k_z$ dependence [cf.\ the lowest four levels in Fig.~\ref{Fig:static_Yp01}(c)]. On the
other hand, the levels with higher energies exhibit $k_z$-dependent band structure, indicating
that the corresponding single-particle wave functions extend spatially outside the slabs.
Those energy levels exhibit a small band gap at $k_z$\,$=$\,0
and $\pm\pi/a$ of the order of tens to hundreds of keV, depending on a pair of the bands
\add{(cf.\ Fig.~\ref{Fig:bandgaps_Yp01})}.
Figure~\ref{Fig:static_Yp01}(d) depicts the parabolic $k_\parallel$ dependence of
$\varepsilon_{\alpha\bs{k}}^{({\rm n})}$, showing both occupied (solid lines)
and unoccupied (dashed lines) states.

We shall now turn to the results of dynamic calculations. In Figs.~\ref{Fig:a_p_Yp01-03}(a),
\ref{Fig:a_p_Yp01-03}(b), and \ref{Fig:a_p_Yp01-03}(c), we show, respectively, acceleration
of protons, $a_{\rm p}(t)$, time derivative of the total linear momentum, $\dot{P}_{\rm tot}(t)$,
and time derivative of the proton linear momentum, $\dot{P}_{\rm p}(t)$, as a function of time.
The end time for switching-on the external force (2,000\,fm/$c$ for the neutron-dripped systems),
is indicated by a vertical line. From the figure, we find again that the slab shows constant acceleration
motion after $t$\,$=$\,2,000\,fm/$c$, especially for the $Y_{\rm p}$\,$=$\,0.3 and 0.2 cases. For the
$Y_{\rm p}$\,$=$\,0.1 case, $a_{\rm p}(t)$ and $\dot{P}_{\rm p}(t)$ shown in Figs.~\ref{Fig:a_p_Yp01-03}(a)
and \ref{Fig:a_p_Yp01-03}(c), respectively, are not precisely constant even when the external force
is constant. It is caused by the presence of dripped neutrons, which fluctuates due to the acceleration
motion of the slab, and some part of the linear momentum is transferred between protons and
dripped neutrons. Since the fluctuation is small, we can quantitatively analyze the collective
masses for the $Y_{\rm p}$\,$=$\,0.1 case as well.

In Fig.~\ref{Fig:Mcoll_Yp01-03}, we show the collective masses (per unit area) as a function
of time for the $Y_{\rm p}$\,$=$\,0.3, 0.2, and 0.1 cases. In Figs.~\ref{Fig:Mcoll_Yp01-03}(a)
and \ref{Fig:Mcoll_Yp01-03}(b), respectively, the collective mass of the slab, $M_{\rm slab}(t)$,
and that of protons, $M_{\rm p}(t)$, are shown. [The panel (c) will be discussed in the next section.]

In order to analyze the results, we here introduce a simple reference for comparison.
A naive estimate of the mass of the slab may be given by the total mass of ``energetically
bound'' nucleons (per unit area) 
\begin{equation}
M_{\rm e.b.} = \sum_q m_q N_q^{\rm e.b.},
\end{equation}
where a simplistic distinction is made with regard to the kinetic energy in the $z$ direction.
Namely, the number of energetically bound nucleons per unit area, $N_q^{\rm e.b.}$,
is evaluated by integrating the density of nucleons within the nuclear potential (i.e.,
$e_{\alpha\bs{k}}^{(q)}$\,$\le$\,$U_0^{(q)}$) over the unit cell \cite{Kashiwaba(2019)}:
\begin{eqnarray}
N_q^{\rm e.b.}
&\equiv& \frac{1}{N_{k_z}}\sum_{\alpha,k_z}\int \frac{k_\parallel}{\pi}
\theta(\mu_q-\varepsilon_{\alpha\bs{k}}^{(q)})
\theta(U_0^{(q)}-e_{\alpha\bs{k}}^{(q)})\,{\rm d}k_\parallel, \nonumber\\[-2.5mm]
\label{Eq:def_N_local}
\end{eqnarray}
where the normalization condition (\ref{Eq:normalization}) was used.
Note that the single-particle energies here are defined by
\begin{equation}
e_{\alpha\bs{k}}^{(q)} \,\equiv\, \varepsilon_{\alpha\bs{k}}^{(q)}\,
-\frac{1}{a}\int_0^a u_{\alpha\bs{k}}^{(q)*}(z)\frac{\hbar^2k_\parallel^2}{2m_q^\oplus(z)}u_{\alpha\bs{k}}^{(q)}(z)\,{\rm d}z,
\label{Eq:e_z}
\end{equation}
removing the kinetic energy (per unit area) associated with the motion parallel
to the slabs (along $x$ and $y$ directions), in accordance with a prescription
proposed in Ref.~\cite{Kashiwaba(2019)}. There hold $N_{\rm n}^{\rm e.b.}
$\,$\le$\,$N_{\rm n}$ and $N_{\rm p}^{\rm e.b.}$\,$=$\,$N_{\rm p}$
in the present case. We note that those nucleons are considered to be bound
in $z$ direction, while they are essentially free in $x$ and $y$ directions parallel
to the slabs. Using the same argument, an estimate of the number of ``free'' or
``energetically unbound'' neutrons $N_q^{\rm f}$ may be given by an equation
similar to Eq.~(\ref{Eq:def_N_local}) with the theta function $\theta(e_{\alpha\bs{k}}^{(q)}-U_0^{(q)})$
or $N_q^{\rm f}=N_q - N_q^{\rm e.b.}$. This defines also the number density
$n_{\rm n}^{\rm f}$\,$=$\,$N_{\rm n}^{\rm f}/a$ of the ``free'' neutrons.

In Fig.~\ref{Fig:Mcoll_Yp01-03}(a) plotted is the ratio of the dynamically obtained
collective mass of the slab, $M_{\rm slab}$, to the reference mass of the ``energetically
bound'' nucleons, $M_{\rm e.b}$. We find intriguing behavior distinct from
the isolated slabs discussed in Sec.~\ref{Sec:isolated}. As shown in Fig.~\ref{Fig:Mcoll_Yp01-03}(a),
the extracted collective mass of the slab is noticeably reduced by more than
10\% and 35\% for the $Y_{\rm p}$\,$=$\,0.2 and 0.1 cases, respectively.
Fig.~\ref{Fig:Mcoll_Yp01-03}(b) shows the collective mass of protons, divided
by the total mass of protons per unit area. We find from Fig.~\ref{Fig:Mcoll_Yp01-03}(b)
that the collective mass of protons is also reduced, although by a smaller amount.
In the next section, we shall explore the cause of the reduction of the collective
masses, observed in the neutron-dripped systems.

\begin{table}[t]
\caption{
The extracted collective masses for the systems of slabs immersed in dripped
neutrons with proton fractions $Y_{\rm p}$\,$=$\,0.3, 0.2, and 0.1 at baryon
density $\bar{n}_{\rm b}$\,$=$\,0.04\,fm$^{-3}$. The collective mass of
protons, $M_{\rm p}$, is shown in the second and the third columns relative to:
$m_{\rm p}N_{\rm p}$, the total mass of protons, and $m_{\rm p,bg}^\oplus
N_{\rm p}$, the total microscopic effective mass of protons in background
neutron density, respectively. The collective mass of a slab, $M_{\rm slab}$,
is shown in the forth and the fifth columns relative to: $M_{\rm e.b.}$\,$=$\,$
m_{\rm p}N_{\rm p}+m_{\rm n}N_{\rm n}^{\rm e.b.}$, the total mass
of ``energetically bound'' nucleons, and $M_{\rm e.b.}^\oplus$\,$=$\,$m_{\rm p,bg}^\oplus
N_{\rm p}+m_{\rm n,bg}^\oplus N_{\rm n}^{\rm e.b.}$, the total
microscopic effective mass of energetically bound nucleons, respectively
[see Eq.~\eqref{Eq:def_N_local} for the definition of $N_q^{\rm e.b.}$].
$M_{\rm slab}$ and $M_{\rm p}$ are mean values averaged over a time interval
from 2,000\,fm/$c$ to 4,000\,fm/$c$, during which the external force is constant.
}\vspace{-1mm}
\label{Table:Masses}
\begin{center}
\begin{tabular*}{\columnwidth}{@{\extracolsep{\fill}}ccccc}
\hline\hline
$Y_{\rm p}$ &
$M_{\rm p}/m_{\rm p}N_{\rm p}$ &
$M_{\rm p}/m_{\rm p,bg}^\oplus N_{\rm p}$ &
$M_{\rm slab}/M_{\rm e.b.}$ &
$M_{\rm slab}/M_{\rm e.b.}^\oplus$ \\
\hline
0.3 & 1.000 & 1.000 & 0.997 & 0.997 \\
0.2 & 0.989 & 1.000 & 0.888 & 0.910 \\
0.1 & 0.970 & 1.001 & 0.655 & 0.703 \\
\hline\hline
\end{tabular*}\vspace{-3mm}
\end{center}
\end{table}

\section{Discussion}\label{Sec:Discussion}

\subsection{The collective mass of protons}\label{Sec:M_p}

Let us first discuss the reason why the collective mass of protons is reduced by a few percent
as compared to the seemingly well-defined total mass of protons, $m_{\rm p}N_{\rm p}$;
see Fig.~\ref{Fig:Mcoll_Yp01-03}(b) and the second column of Table~\ref{Table:Masses}.
Actually, we find that the reduction is caused by the density-dependent microscopic effective
mass. Namely, the extracted collective mass is very close to the total effective mass of protons,
\begin{equation}
M_{\rm p} \simeq m_{\rm p,bg}^\oplus N_{\rm p},
\label{Eq:M_p_eff}
\end{equation}
where $m_{\rm p,bg}^\oplus$\,$\equiv$\,$m_{\rm p}^\oplus[n_{\rm n}^{\rm bg}]$ is the
microscopic proton effective mass at the background neutron density, $n_{\rm n}^{\rm bg}$;
see the third column of Table~\ref{Table:Masses}. It is not obvious since the microscopic
proton effective mass should be much smaller inside the slabs, where nucleon density is
higher than the outside. Our time-dependent simulations suggest that protons inside
a nuclear cluster behave as if they have the microscopic effective mass in uniform
background neutron matter.

The source of reduction can be inferred by analyzing the continuity equations within
the Skyrme TDDFT approach. Note that the discussion here is applicable to any shape
of nuclear cluster (not necessary a slab), and the equations below are written in a 3D
coordinate system. As usual, the total momentum density satisfies the continuity equation,
\begin{equation}
\frac{\partial\rho(\bs{r},t)}{\partial t} + \hbar\bs{\nabla\cdot}\bs{j}(\bs{r},t) = 0,
\label{Eq:continuity_tot}
\end{equation}
where $\rho(\bs{r},t)$\,$=$\,$\sum_qm_qn_q(\bs{r},t)$ is the nucleon mass density. On the other
hand, each component (for neutrons or protons) of the momentum densities does not satisfy the
ordinary continuity equation, due to mutual interactions between the two species. By carefully
taking the time derivative of the mass density, and using the explicit form of the TDKS equations,
one can derive the following continuity equations for neutrons ($q$\,$=$\,n) or protons ($q$\,$=$\,p)
\cite{Engel(1975),Chamel(2019)}:
\begin{equation}
\frac{\partial \rho_q(\bs{r},t)}{\partial t} + \hbar\bs{\nabla\cdot}\bs{p}_q(\bs{r},t) = 0,
\label{Eq:continuity_q}
\end{equation}
where the modified momentum density is defined by
\begin{equation}
\bs{p}_q(\bs{r},t) \equiv
\frac{m_q}{m_q^\oplus(\bs{r},t)}\bs{j}_q(\bs{r},t) + \frac{m_q}{\hbar^2}n_q(\bs{r},t)\bs{I}_q(\bs{r},t),
\end{equation}
or more explicitly,
\begin{eqnarray}
\bs{p}_q(\bs{r},t)
= \bs{j}_q(\bs{r},t) &+& q\frac{2m_q}{\hbar^2}\bigl(C_0^\tau-C_1^\tau\bigr)\,n_{\rm n}(\bs{r},t)n_{\rm p}(\bs{r},t) \nonumber\\
&&\times\,
\biggl( \frac{\bs{j}_{\rm p}(\bs{r},t)}{n_{\rm p}(\bs{r},t)} - \frac{\bs{j}_{\rm n}(\bs{r},t)}{n_{\rm n}(\bs{r},t)} \biggr),
\label{Eq:j_tilde}
\end{eqnarray}
where $q$\,$=$\,$+1$ for protons and $q$\,$=$\,$-1$ for neutrons. The additional
term stems from mutual friction which is associated with a velocity lag between the
different species [cf.\ the second term in Eq.~(\ref{Eq:j_tilde})]. It should be
remarked that it is closely related to the so-called entrainment matrix and the isovector
effective mass and, thus, the enhancement factor of the energy-weighted sum rule,
as shown in Ref.~\cite{Chamel(2019)}. Neglecting the small difference between
neutron and proton masses, the additional term cancels out when we take the sum,
giving rise to the usual total momentum density, i.e. $\sum_q\bs{p}_q(\bs{r},t)
$\,$=$\,$\bs{j}(\bs{r},t)$. We have confirmed that a modified proton collective
mass defined by $\widetilde{M}_{\rm p}(t)\equiv(1/a_{\rm p}(t)){\rm d}\widetilde
{P}_{\rm p}(t)/{\rm d}t$, with $\widetilde{P}_{\rm p}(t)$\,$\equiv$\,$\hbar
\int_0^a p_{z,{\rm p}}(z,t)\,{\rm d}z$, coincides with the total bare proton mass,
as shown in Fig.~\ref{Fig:Mcoll_Yp01-03}(c). It is because of the fact that the modified
momentum density does satisfy the continuity equation (\ref{Eq:continuity_q}).
This observation suggests that the reduction of the proton collective mass observed
in Fig.~\ref{Fig:Mcoll_Yp01-03}(b) and the second column of Table~\ref{Table:Masses}
is caused by the velocity difference between protons and background neutrons.

\subsection{Counterflow of ``anti-entrained'' neutrons}\label{Sec:jz}

Next, we discuss the sizable reduction of the collective mass of the slab, $M_{\rm slab}$,
observed in Fig.~\ref{Fig:Mcoll_Yp01-03}(a). The discussion in Sec.~\ref{Sec:M_p}
indicates that the microscopic effective mass plays a certain role in the reduction of the
collective masses. Naively, one may assume that all the ``energetically bound'' neutrons
also move with their microscopic effective mass in the background neutron density,
as found for protons as shown in Sec.~\ref{Sec:M_p}. With this assumption, the collective mass
of the slab may be estimated by $M_{\rm e.b.}^\oplus \equiv m_{\rm p,bg}^\oplus N_{\rm p}+ m_{\rm n,bg}^\oplus
N_{\rm n}^{\rm e.b.}$ with $m_{\rm n,bg}^\oplus$\,$=$\,$m_{\rm n}[n_{\rm n}^{\rm bg}]$.
In the fifth column of Table~\ref{Table:Masses}, the ratio, $M_{\rm slab}/
M_{\rm e.b.}^\oplus$, is displayed. If this hypothesis works, the ratio
should be unity for all $Y_{\rm p}$. Apparently, the ratio becomes substantially
smaller than unity especially for $Y_{\rm p}$\,$=$\,0.1, meaning that the observed
reduction of the collective mass of the slab cannot be explained solely by the microscopic
effective mass. It indicates that there exists some other mechanism that reduces
the collective mass of the slab immersed in a sea of dripped neutrons, e.g., some
effects of hydrodynamic origin (cf.\ Ref.~\cite{Martin(2016)}) and/or those
associated with the Bragg scattering of dripped neutrons by the periodic potential,
i.e., band structure effects.

\begin{figure} [t]
\includegraphics[width=\columnwidth]{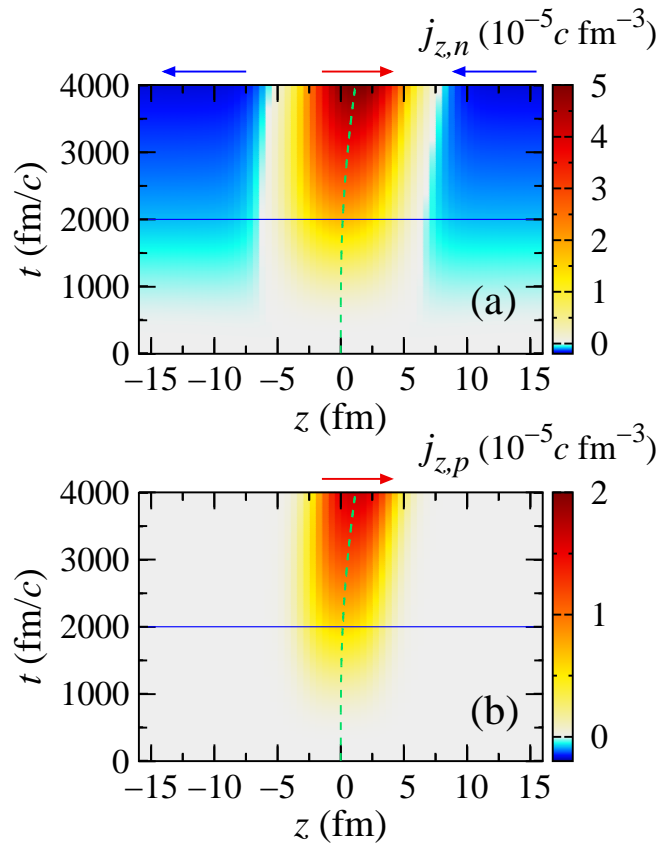}\vspace{-1mm}
\caption{
Results of TDDFT calculations for the $Y_{\rm p}$\,$=$\,0.1 case at baryon density
$\bar{n}_{\rm b}$\,$=$\,0.04\,fm$^{-3}$. In panels (a) and (b), the current densities
of neutrons and protons, $j_{z,{\rm n}}$ and $j_{z,{\rm p}}$, are shown in the $z$-$t$
plane, respectively, in the units of $c$\:fm$^{-3}$. The green dashed curve represents the
trajectory of the center-of-mass position of protons, $Z_{\rm p}(t)$. The blue horizontal line
indicates the time up to which the external force is turned on. The arrows at the top of those
panels guide the directions of the currents.
}\vspace{-3mm}
\label{Fig:jz_Yp01}
\end{figure}

A great advantage of using the real-time approach is that it allows us to gain intuitive
understanding of complex many-body dynamics. Indeed, from the time-dependent
response of the system to the external force, we find the cause of the reduction.
In Fig.~\ref{Fig:jz_Yp01}, we show the time evolution of the current density (in units
of $c$\:fm$^{-3}$, i.e., the momentum density is multiplied by the factor $\hbar/m_q$)
in the $z$-$t$ plane for the $Y_{\rm p}$\,$=$\,0.1 case. In Figs.~\ref{Fig:jz_Yp01}(a)
and \ref{Fig:jz_Yp01}(b), the current densities of neutrons and protons, $j_{z,{\rm n}}$
and $j_{z,{\rm p}}$, are shown, respectively. In the figure, a trajectory of the center-of-mass
position of protons, $Z_{\rm p}(t)$, is also indicated by a dashed curve. The external force is
oriented towards the positive-$z$ direction, which is smoothly turned on from 0 to 2,000\,fm/$c$,
and the end time is indicated by a horizontal line.

In Figs.~\ref{Fig:jz_Yp01}(a) and \ref{Fig:jz_Yp01}(b), the yellow to red colored
area (around $z$\,$=$\,0, close to the slab) corresponds to the region where the
current density is positive, i.e., both neutrons and protons around the slab flow towards
the same direction as the external force, as expected. The magnitude of the current
density increases with time, and one can indeed see a slight movement of the slab towards
positive-$z$ direction (see the dashed line). As is evident from Fig.~\ref{Fig:jz_Yp01}(a), however,
the neutron current density becomes actually negative outside the slab (indicated by cyan
to blue colored area). It means that the dripped neutrons outside the slab flows towards
the direction opposite to the external force. Since the ``counterflow'' of neutrons reduces
the magnitude as well as the rate of increase of the total linear momentum, it results in
the reduction of the collective mass of the slab [cf.\ Eq.~(\ref{Eq:M_slab})]. The
counterflow indicates that a part of dripped neutrons is coupled to the periodic potential
of the slabs, but in a way opposite to the original picture of the entrainment effect. Namely,
in the rest frame of the slabs, for instance, the band structure acts as lubricant for the dripped
neutrons. We may call it an ``anti-entrainment effect.'' The emergence of the neutron
counterflow could be qualitatively attributed to a band structure effect, as discussed in Sec.~\ref{Sec:comparison}.

\subsection{Conduction neutron density and macroscopic effective mass}\label{Sec:comparison}

While we observe the anti-entrainment effect in our time-dependent simulations,
a static formalism has been utilized in the literature \cite{Chamel(2017),Kashiwaba(2019)}.
In the following, we demonstrate that the results of our dynamic calculations are consistent
with the static approach.

\subsubsection{Static approach}\label{Sec:static}

In the static approach \cite{Chamel(2017),Kashiwaba(2019)}, the entrainment
effect has been discussed in terms of the mobility coefficients, $\mathcal{K}_{\mu\nu}$,
the conduction neutron number density, $n_{\rm n}^{\rm c}$, and the corresponding
``macroscopic'' effective mass of neutrons, $m_{\rm n}^\star$. The mobility coefficients
are defined by
\begin{equation}
\mathcal{K}_{\mu\nu}^{({\rm n})}
= 2\sum_{\alpha\bs{k}}^{\rm occ.}
\Bigl(m_{{\rm n},\alpha\bs{k}}^{\star\,-1}\Bigr)_{\mu\nu},
\label{Eq:K_munu}
\end{equation}
where 
\begin{equation}
\Bigl(m_{{\rm n},\alpha\bs{k}}^{\star\,-1}\Bigr)_{\mu\nu}
= \frac{1}{\hbar^2}\frac{\partial^2\varepsilon_{\alpha\bs{k}}^{({\rm n})}}{\partial k_\mu\partial k_\nu}.
\label{Eq:def_mstar_tensor}
\end{equation}
is the inverse of the so-called macroscopic effective mass tensor. It is noted here that
for a uniform neutron matter with density $n$ the macroscopic effective mass tensor
coincides with the \textit{microscopic} effective mass $m_{\rm n}^\oplus=m_{\rm n}^\oplus[n]$, i.e.
$\bigl(m^{\star\,-1}_{{\rm n},\alpha\bs{k}}\bigr)_{\mu\nu}=\delta_{\mu\nu}/m_{\rm n}^\oplus$
and $\mathcal{K}_{\mu\nu}^{({\rm n})}=\delta_{\mu\nu}n_{\rm n}/m_{\rm n}^\oplus$.

Since energies of bound orbitals do not have $\bs{k}$ dependence, their
macroscopic effective mass tensor is divergently large, that is, $\bigl(m_{{\rm n},
\alpha\bs{k}}^{\star\,-1}\bigr)_{\mu\nu}$\,$\approx$\,0. In other words
only such neutrons that belong to an energy band with $\bs{k}$ dependence can
contribute to the sum in the mobility coefficients \eqref{Eq:K_munu}. As it quantifies
neutrons that can contribute to the conduction, it allows us to define the conduction
neutron number density. Namely, for the slab phase under study, we may define the
conduction neutron number density as
\begin{equation}
n_{\rm n}^{\rm c} \equiv m_{\rm n,bg}^\oplus\mathcal{K}_{zz}^{({\rm n})},
\label{Eq:n_c_assumption}
\end{equation}
by normalizing the mobility coefficient with the effective mass $m_{\rm n,bg}^\oplus$
expected for neutrons in the uniform background. The mobility coefficient for $z$ direction
normal to the slabs reads:
\begin{equation}
\mathcal{K}_{zz}^{({\rm n})} = \frac{1}{\pi L}\sum_{\alpha,k_z}\int k_\parallel
\,\Bigl(m_{{\rm n},\alpha\bs{k}}^{\star\,-1}\Bigr)_{zz}
\theta(\mu_{\rm n}-\varepsilon_{\alpha\bs{k}}^{({\rm n})})\,{\rm d}k_\parallel.
\end{equation}

The \textit{macroscopic} effective mass of neutrons, $m_{\rm n}^\star$,
can also be defined using the mobility coefficient. Namely, the macroscopic
effective mass is introduced regarding that all the ``free'' neutrons with density
$n_{\rm n}^{\rm f}$ contribute to the conduction, but having an average
effective mass, $m_{\rm n}^\star$, leading to the following definition:
\begin{equation}
{n_{\rm n}^{\rm f}} \equiv m_{\rm n}^{\star}K_{zz}^{({\rm n})}.
\label{Eq:mstar_definition_1}
\end{equation}
Combining Eqs.~\eqref{Eq:n_c_assumption} and \eqref{Eq:mstar_definition_1},
one finds a relation,
\begin{equation}
\frac{m_{\rm n}^\star}{m_{\rm n,bg}^\oplus} = \frac{n_{\rm n}^{\rm f}}{n_{\rm n}^{\rm c}}.
\label{Eq:def_mstar}
\end{equation}
The conduction neutron number density (\ref{Eq:n_c_assumption}) and
the corresponding macroscopic effective mass (\ref{Eq:def_mstar}) obtained
with the static approach are presented in the third and the fourth columns of
Table~\ref{Table:comparison}.

\subsubsection{Time-dependent approach}

As described in the previous section, the static approach requires an elaborated
analysis of complex band structure. The complexity increases when one considers
crustal phases with higher dimensions (e.g., rods and spheres) \cite{Chamel(2017)}.
Here we show that we can easily estimate the conduction neutron number density
from the collective masses obtained within the time-dependent approach. It offers
an alternative and intuitive way of analyzing the band structure effect, which can
be applied also for higher dimensions.

Since we have separately extracted the collective mass of a slab and that of
protons, we can naturally define the collective mass of neutrons which are
``effectively bound'' to the slab as follows:
\begin{equation}
M_{\rm n}^{\rm eff.bound} = M_{\rm slab} - M_{\rm p}.
\label{Eq:M_n_eff.bound}
\end{equation}
Here, ``effectively bound'' means both bound neutrons inside the slab
and dripped neutrons part of which are entrained to the slab because of
the band structure effect. We then quantify the number of effectively
bound neutrons via the relation,
\begin{equation}
M_{\rm n}^{\rm eff.bound} = m_{\rm n,bg}^\oplus N_{\rm n}^{\rm eff.bound},
\label{Eq:M_n_eff.bound_2}
\end{equation}
which defines the average number density of the effectively-bound neutrons,
\begin{equation}
n_{\rm n}^{\rm eff.bound}
\;=\; \frac{N_{\rm n}^{\rm eff.bound}}{a}
\;=\; \frac{M_{\rm n}^{\rm eff.bound}}{m_{\rm n,bg}^\oplus a}.
\end{equation}
Assuming that the conduction neutrons are the counterpart of the effectively-bound
neutrons, the conduction neutron number density can be defined as follows:
\begin{equation}
n_{\rm n}^{\rm c} \equiv \bar{n}_{\rm n}-n_{\rm n}^{\rm eff.bound}.
\label{Eq:n_c_dynamic}
\end{equation}
The conduction neutron number density (\ref{Eq:n_c_dynamic}) obtained from the
time-dependent approach are presented in the fifth columns of Table~\ref{Table:comparison}.

\subsubsection{Comparison}

\begin{table}[t]
\caption{
Results of static and dynamic self-consistent band calculations for the systems
of slabs immersed in dripped neutrons with proton fractions $Y_{\rm p}$\,$=$\,0.3,
0.2, and 0.1 at baryon density $\bar{n}_{\rm b}$\,$=$\,0.04\,fm$^{-3}$.
The number density of dripped (or ``free'') neutrons, $n_{\rm n}^{\rm f}$
[=\,$\bar{n}_{\rm n}-n_{\rm n}^{\rm e.b.}$, where $n_{\rm n}
^{\rm e.b.}$\,$=$\,$N_{\rm n}^{\rm e.b.}/a$; cf.\ Eq.~(\ref{Eq:def_N_local})]
is shown in the second column. The conduction neutron number density relative
to the average neutron number density $n_{\rm n}^{\rm c}/\bar{n}_{\rm n}$,
and the corresponding macroscopic effective mass relative to the bare neutron mass
$m_{\rm n}^\star/m_{\rm n}$ (\ref{Eq:def_mstar}) are listed in the third to the
fifth columns. The third and the forth columns show those obtained with the static
approach [Eq.~(\ref{Eq:n_c_assumption})], while those obtained with the dynamic
one [Eq.~(\ref{Eq:n_c_dynamic})] are shown in the fifth column.
}\vspace{-1mm}
\label{Table:comparison}
\begin{center}
\begin{tabular*}{\columnwidth}{@{\extracolsep{\fill}}cccccccc}
\hline\hline
\multirow{2}{*}{$Y_{\rm p}$} & {\;\;\;} &
\multirow{2}{*}{$n_{\rm n}^{\rm f}/\bar{n}_{\rm n}$\;} &
\multicolumn{2}{c}{Static} & {\;\;\;} &
\multicolumn{1}{c}{Dynamic} & {} \\\cline{4-5}\cline{7-7}
{} & {} & 
{} &
\;$n_{\rm n}^{\rm c}/\bar{n}_{\rm n}$\; &
\;$m_{\rm n}^\star/m_{\rm n}$\; & {} &
\;$n_{\rm n}^{\rm c}/\bar{n}_{\rm n}$\; & {} \\
\hline
0.3 & {} & 
2.09\,$\times$\,10$^{-4}$ &
0.005 & 0.040 & {} & 0.005  & {}\\
0.2 & {} & 
0.127 &
0.256 & 0.496 & {} & 0.229 & {}\\
0.1 & {} & 
0.362 &
0.630 & 0.574 & {} & 0.586 & {}\\
\hline\hline
\end{tabular*}\vspace{-3mm}
\end{center}
\end{table}

In Table~\ref{Table:comparison}, the results of static and dynamic
calculations for $Y_{\rm p}$\,$=$\,0.3, 0.2, and 0.1 are summarized.
In the second column, the ``free'' neutron density [$n_{\rm n}^{\rm f}
= \bar{n}_{\rm n}-n_{\rm n}^{\rm e.b.}$, where $n_{\rm n}^{\rm e.b.}
$\,$=$\,$N_{\rm n}^{\rm e.b.}/a$; cf.\ Eq.~(\ref{Eq:def_N_local})]
relative to the mean neutron number density $\bar{n}_{\rm n}$ is shown,
as a reference density. In the third and the fifth columns, the conduction
neutron number densities obtained with the static and the dynamic approaches
[i.e., Eqs.~\eqref{Eq:n_c_assumption} and \eqref{Eq:n_c_dynamic}],
are shown, respectively.

From Table~\ref{Table:comparison}, we find that the results of the static
and dynamic calculations reasonably agree with each other. The results indicate
that the conduction neutron number density in the static approach tends to be
slightly larger than that in the dynamic one (cf.\ the third and the fifth columns
of Table~\ref{Table:comparison}). Note that we made assumptions for the effective
mass of neutrons [Eqs.~\eqref{Eq:n_c_assumption}, \eqref{Eq:M_n_eff.bound_2},
and \eqref{Eq:n_c_dynamic}]. If the microscopic effective mass in the slab
region contributes, where it takes a smaller value than that outside of the slabs,
$n_{\rm n}^{\rm c}$ would become smaller than current estimates. A possible
prescription may be to take an average, $\bar{m}_{\rm n}^\oplus=\frac{1}{N_{\rm n}}
\int m_{\rm n}^\oplus(z)n_{\rm n}(z){\rm d}z$, which we shall investigate
for realistic proton fractions in future work.

In both static and dynamic calculations, the conduction neutron number density is found
to be substantially larger than the ``free'' neutron number density. It means that neutrons
are not actually entrained, but rather mobilized via the Bragg scattering by the periodic potential,
at least in the slab phase of nuclear matter. This conclusion is consistent with the work of
Ref.~\cite{Kashiwaba(2019)} where the static approach with a BCPM EDF \cite{Baldo(2013)}
was employed. Note that the entrainment effect which is discussed for 3D lattices of spherical
nuclear clusters \cite{Chamel(2017)} corresponds to a case of $n_{\rm n}^{\rm c}
< n_{\rm n}^{\rm f}$ and $m_{\rm n}^\star > m_{\rm n}^\oplus$, which implies that
only a part of the ``free'' neutrons contribute to the conduction while the remaining part is
bound to the lattice of clusters. Based on the newly-developed time-dependent approach,
we have confirmed the conclusion of Ref.~\cite{Kashiwaba(2019)} and revealed the
emergence of the neutron counterflow (as shown in Sec.~\ref{Sec:jz}), which leads
to a new interpretation as an ``anti-entrainment'' effect. Note that the static and
time-dependent approaches focus on different aspects of the entrainment phenomenon.
In the static approach the amount of neutrons that can actually conduct is analyzed.
In contrast, in the time-dependent approach we focus on the complementary aspect,
that is, the amount of neutrons which are effectively bound to the slabs that can not
conduct. The agreement between the two approaches justify our assumption
\eqref{Eq:n_c_dynamic} and the interpretation.

Qualitatively, the observed anti-entrainment could be attributed to a band structure effect.
Namely, the rate of change of the group velocity of dripped neutrons can be either positive
or negative depending on the sign of the macroscopic effective mass tensor [cf.\ Eq.~(\ref{Eq:def_mstar_tensor})].
It means that if the effective mass tensor is negative (i.e., the energy band is convex
\add{downward} as a function of the Bloch wave number), the corresponding state is accelerated
towards the direction opposite to the external force. It stems from the Bragg scattering which
causes exchange of momentum between the periodic structure and the dripped neutrons.

\begin{figure} [t]
\includegraphics[width=\columnwidth]{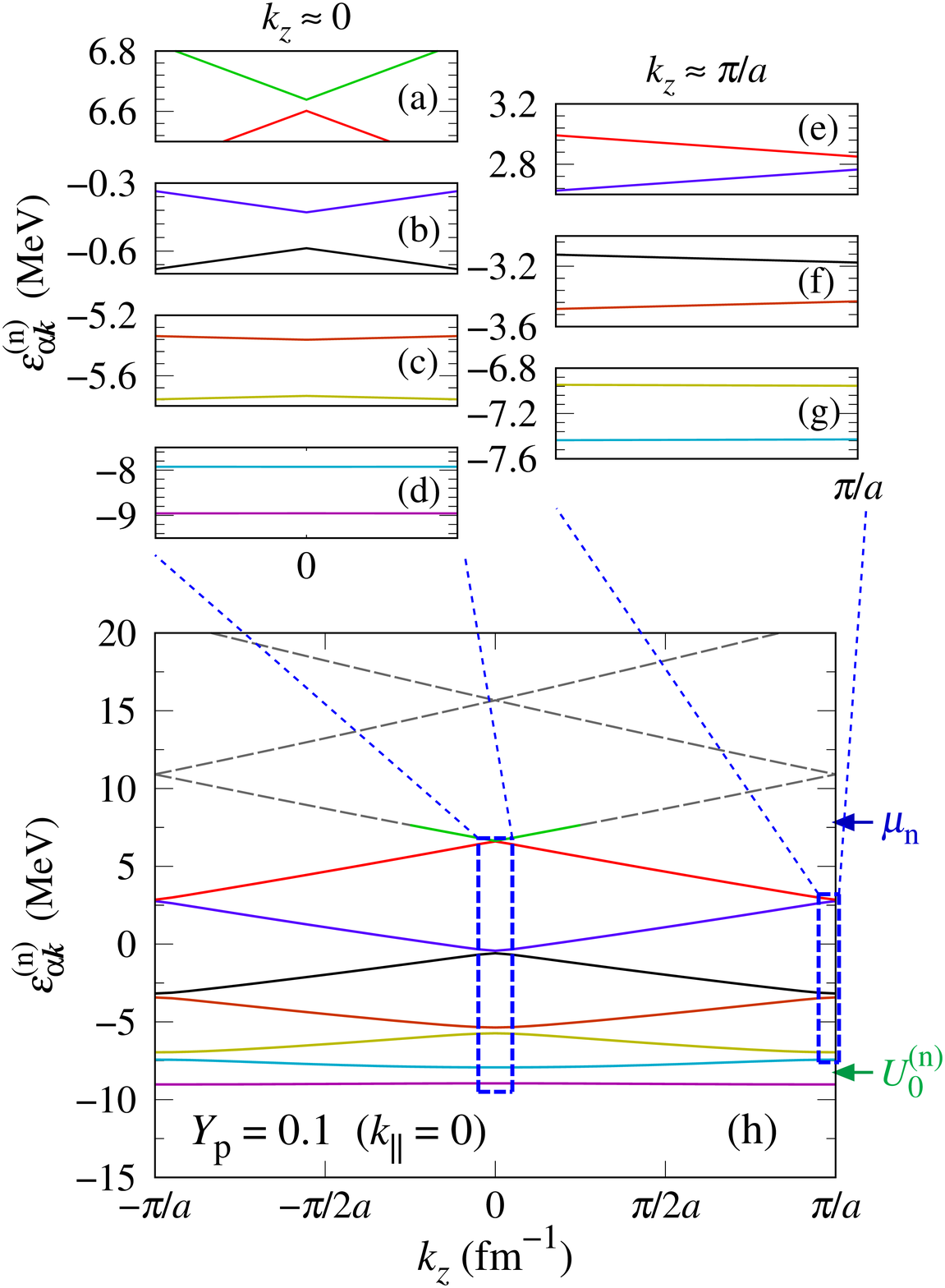}\vspace{-1mm}
\caption{
\add{
Results of the static self-consistent band calculation for the system of slabs
immersed in dripped neutrons with proton fractions $Y_{\rm p}$\,$=$\,0.1 at baryon
density $\bar{n}_{\rm b}$\,$=$\,0.04\,fm$^{-3}$. In panel (h),
$\varepsilon_{\alpha\bs{k}}^{({\rm n})}$ for $k_\parallel=0$ are shown as a function
of the Bloch wave number $k_z$ within the first Brillouin zone, $-\pi/a$\,$\le$\,$k_z$\,$\le$\,$\pi/a$
[same as Fig.~\ref{Fig:static_Yp01}(c), but for $-15$\,MeV\,$\le$\,$\varepsilon_{\alpha\bs{k}}^{({\rm n})}$\,$\le$20\,MeV].
In panels (a)--(d), magnified plots around the center of the first Brillouin zone, $-0.05\pi/a\le k_z\le+0.05\pi/a$, are shown.
In panels (e)--(g), those around the edge of the first Brillouine zone, $0.95\pi/a\le
k_z\le\pi/a$, are shown. The corresponding resions are indicated by blue dashed boxes
in panel (h).
}
}\vspace{-3mm}
\label{Fig:bandgaps_Yp01}
\end{figure}

\add{
In Fig.~\ref{Fig:bandgaps_Yp01}(h), we again show single-particle energies
$\varepsilon_{\alpha\bs{k}}^{({\rm n})}$ for $k_\parallel=0$ as a function of the Bloch
wave number $k_z$ within the first Brillouin zone, $-\pi/a$\,$\le$\,$k_z$\,$\le$\,$\pi/a$
[same as Fig.~\ref{Fig:static_Yp01}(c), but for $-15$\,MeV\,$\le$\,$\varepsilon_{\alpha\bs{k}}^{({\rm n})}$\,$\le$20\,MeV].
The orbitals with $\varepsilon_{\alpha\bs{k}}^{({\rm n})}>U_0^{(\mathrm{n})}$ exhibit
a band structure and, thus, contribute to the macroscopic effective mass. The contributions
to the macroscopic effective mass come mainly from the center and the edge of the first
Brillouin zone, where the second derivative of energy bands become substantial.
For better visibility, magnified plots around the center and
the edge of the first Brillouin zone are shown in panels (a)--(d) and (e)--(g), respectively.
From the plots, one can clearly see that there are pairs of bands, having opposite signs
of the second derivative, around the center and the edge of the first Brillouin zone.
It explains why we observe the neutron counter flow which flows opposite to the external
force, as a part of dripped neutrons has a negative macroscopic effective mass tensor.
}

\section{Summary and prospect}\label{Sec:Summary}

In the inner crust of neutron stars crystalline structure is formed in competition
with the Coulomb and surface energies, which coexists with dripped neutrons.
The dripped neutrons are affected by the periodic structure via the Bragg scattering.
A proper quantum mechanical method to describe a system with a periodic potential
is the band theory of solids, where the periodicity is incorporated with the Bloch
boundary condition. Although the band theory has been a standard tool in solid-state
physics, its application to nuclear systems is rather new, and the first fully self-consistent
static band theory calculations were reported only in 2019 \cite{Kashiwaba(2019)}.

In the present paper, we have developed a time-dependent extension of
the fully self-consistent band theory for nuclear systems to investigate dynamic
properties of the inner crust of neutron stars, combining methods from nuclear
and solid-state physics. We employ the time-dependent density functional theory
(TDDFT) formalism for nuclear systems, incorporated with the Bloch boundary
condition. In this work, we have applied the formalism to the slab phase of
nuclear matter, with a Skyrme-type energy density functional, neglecting
the spin-orbit coupling and pairing correlations.

From a dynamic response of the system to an external force, we have extracted the collective
mass of a slab, and that of protons inside the slab, from which we could obtain the conduction
neutron number density. The latter is a key quantity related to the ``macroscopic'' effective
mass---an effective mass of dripped neutrons under the influence of the periodic potential
(the so-called entrainment effects)---which enters various macroscopic models of the inner
crust of neutron stars. From the results, we have found that the collective mass of a slab is
lighter than expected from a naive estimation based on the maximum value of the mean-field
potential and the single-particle energies. We have also found that the collective mass of
protons is reduced as well by a few percent from the total bare proton mass. The latter could
be explained by the microscopic effective mass in background neutron density. On the contrary,
the reduction of the collective mass of the slab could not be explained solely by the microscopic
effective mass. From the time evolution of the system, we have found that part of dripped neutrons,
which are expected to be entrained to the periodic potential due to the Bragg scattering, are
actually moving towards the direction opposite to the motion of the slabs. Namely, in the rest
frame of the crust neutron currents shall be enhanced. We call it the ``anti-entrainment'' effect.
As a consequence, the collective mass of the slab is reduced, which, in turn, corresponds
to an enhancement of the conduction neutron number density and
a reduction of the macroscopic effective mass of dripped neutrons. We note that this result
is consistent with the recent self-consistent band calculations of Ref.~\cite{Kashiwaba(2019)},
while our time-dependent analysis sheds new light on the underlying physics that was not clear
in the static approach.

It is interesting to notice that the observed counterflow of dripped neutrons is somewhat similar
to the velocity field predicted by the superfluid hydrodynamics approach \cite{Martin(2016)},
where similar flow pattern was found mainly around nuclear clusters. However, we should note that
they are of completely different origins. In our case, it is originated from the band structure effects which
allow for part of dripped neutrons to have negative effective mass that could make their response opposite
to the external force. On the contrary, in the superfluid hydrodynamics approach \cite{Martin(2016)},
effective mass of a nuclear cluster is quantified by solving equations for the velocity potential $\varphi(\bs{r})$
(or the phase of the superfluid order parameter) under appropriate boundary conditions, which
determines superfluid velocity $\bs{v}_{\rm s}(\bs{r})$\,$\propto$\,$\bs{\nabla}\varphi(\bs{r})$
and density, and hence band structure effects are absent. On one hand, there is an argument
that pairing correlations may hinder band structure effects \cite{Watanabe(2017)}, on the other hand,
the energy gap is actually $\bs{k}$-dependent (i.e., $\Delta \varepsilon_{\alpha\bs{k}}^{({\rm n})}
\equiv \varepsilon_{\alpha+1,\bs{k}}^{({\rm n})}-\varepsilon_{\alpha\bs{k}}^{({\rm n})}$
could be both $\Delta \varepsilon_{\alpha\bs{k}}^{({\rm n})}\ll\Delta_{\rm n}$ and
$\Delta \varepsilon_{\alpha\bs{k}}^{({\rm n})}\gg\Delta_{\rm n}$ for different $\bs{k}$
values, where $\Delta_{\rm n}$ \add{here represents} the average pairing gap, see
also Fig.~\ref{Fig:bandgaps_Yp01}) and the pairing properties may vary significantly
with the Bloch wave vector $\bs{k}$. Thus, it seems premature to draw a conclusion of
pairing effects on the entrainment phenomenon, before a fully self-consistent band theory calculation
including superfluidity is achieved. An important task is to extend our formalism to include superfluidity
based on TDDFT for superfluid systems, the so-called time-dependent superfluid local density approximation
(TDSLDA) (see, e.g., Refs.~\cite{Bulgac(2012),Bulgac(2019)}). In the latter case, superfluid dynamics are
described self-consistently and microscopically on top of quasiparticle energy levels with band structure,
and its extension to finite temperatures is straightforward (see, e.g., Ref.~\cite{Daniel(2021)}).
It will certainly provide us useful information to resolve the debatable situation of the effects of
entrainment in the inner crust of neutron stars, which affects interpretations of astrophysical
phenomena of neutron stars, such as pulsar glitches \cite{Andersson(2012),Chamel(2013)glitch}.

By extending the present formalism to 2D or 3D geometry, much more abundant physics involving
dynamics of both dripped neutrons and nuclear clusters can be examined, in a fully microscopic way.
We mention here that the nuclear pasta phase was already examined within 3D TDDFT (without superfluidity)
\cite{Schuetrumpf(2013),Schuetrumpf(2014),Schuetrumpf(2015)1,Schuetrumpf(2015)2}, and also
dynamics of a vortex-nucleus system were studied within 3D TDSLDA including superfluidity \cite{VNdyn(2016)},
although band structure effects were not taken into account in those works. At low temperatures,
long-wavelength phonons are mainly responsible for the thermodynamic properties of the neutron
star crust. In this regard, various lattice oscillation modes and their coupling to superfluid phonons
have been studied within macroscopic effective theories \cite{Chamel(2013),Kobyakov(2013),
Kobyakov(2016),Durel(2018)}. Within the time-dependent self-consistent band theory, we will
be able, in principle, to examine various oscillation modes of the crustal nuclear matter microscopically
to determine, e.g., share coefficients, phonon velocities, specific heat, and so on. Besides, it has
been expected that the quasi-periodic oscillations, observed in an afterglow of $X$-ray bursts in
some magneters, are associated with torsional oscillations of nuclear lattices in the inner crust of
the neutron stars. Recently, quasiperiodic oscillations, originally observed in the $X$-ray afterglow
of the giant flare of SGR\,1806-20 as well as those recently found by a Bayesian analysis \cite{Miller(2019)},
have been nicely explained by torsional oscillations of the nuclear pasta \cite{Sotani(2019)},
which serves as an evidence of the existence of the pasta phase. It would be of great importance
to provide microscopic inputs to and/or validation of those macroscopic models of quasiperiodic
oscillations. The fully self-consistent time-dependent nuclear band theory proposed here will
pave a new way for developing microphysics-based macroscopic models for diverse phenomena
relevant to the inner crust of neutron stars.

\begin{acknowledgments}
K.S.\ acknowledges Piotr Magierski and Gabriel Wlaz{\l}owski (Warsaw University
of Technology) for an exploratory collaborative work on the dynamic method for
extraction of the collective mass of a nuclear cluster immersed in neutron superfluid
that stimulated this work.
K.S.\ is thankful to K.~Yoshimura for careful reading of the manuscript and providing
valuable comments on the article.
K.S.\ is also grateful for the organization of the INT Program
(INT-19-1a), \textit{Quantum Turbulence: Cold Atoms, Heavy Ions, and Neutron Stars},
for fruitful discussions during the program that were useful to conduct this work.
This work was supported by JSPS KAKENHI Grant Numbers 17K34567 and 20K03945.
\end{acknowledgments}

\appendix

\section{The length and velocity gauges}\label{App:gauges}

Since the velocity gauge may not be a well-known concept in nuclear physics
community, here we provide succinct description on gauge transformations.
Let us consider a time-dependent Schr\"odinger equation (TDSE) for a particle
with mass $m$ and charge $e$ under scalar and vector potentials, $\phi(\bs{r},t)$
and $\bs{A}(\bs{r},t)$, respectively:
\begin{eqnarray}
\zI\hbar\frac{\partial\psi(\bs{r},t)}{\partial t}
&=& \hat{H}\psi(\bs{r},t) \\[1mm]
&=& \biggl[ \frac{1}{2m}\Bigl(\hat{\bs{p}}-\frac{e}{c}\bs{A}(\bs{r},t)\Bigr)^2 + e\phi(\bs{r},t) \biggl]\psi(\bs{r},t).\nonumber\\[-1mm]
\label{Eq:TDSE}
\end{eqnarray}
This equation has a redundant degree of freedom with an arbitrary function
$\chi(\bs{r},t)$. Namely, simultaneous transformations,
\begin{eqnarray}
\bs{A}'(\bs{r},t) &=& \bs{A}(\bs{r},t) +\bs{\nabla}\chi(\bs{r},t),\\[1mm]
\phi'(\bs{r},t) &=& \phi(\bs{r},t)-\frac{1}{c}\frac{\partial\chi(\bs{r},t)}{\partial t},\\[1mm]
\psi'(\bs{r},t) &=& U(\bs{r},t)\psi(\bs{r},t),
\end{eqnarray}
leave the electric field $\bs{E}(\bs{r},t)=-\bs{\nabla}\phi(\bs{r},t)-\frac{1}{c}\partial\bs{A}(\bs{r},t)/\partial t$,
the magnetic field $\bs{B}(\bs{r},t)=\bs{\nabla\times}\bs{A}(\bs{r},t)$, and the TDSE \eqref{Eq:TDSE}
unchanged. Here, $U(\bs{r},t)$ denotes a unitary (gauge) transformation,
\begin{equation}
U(\bs{r},t) = \exp\biggl[-\frac{\zI e}{\hbar c}\chi(\bs{r},t)\biggr].
\label{Eq:Utrans}
\end{equation}
Corresponding to the gauge transformation of the wave function,
time derivative and momentum operators are transformed, respectively, as
\begin{eqnarray}
\partial_t' &\equiv& U(\bs{r},t) \frac{\partial}{\partial t} U^\dagger(\bs{r},t) = \frac{\partial}{\partial t} + \frac{\zI e}{\hbar c}\frac{\partial\chi(\bs{r},t)}{\partial t}, \\[1mm]
\hat{\bs{p}}' &\equiv& U(\bs{r},t) \hat{\bs{p}} U^\dagger(\bs{r},t) = \hat{\bs{p}} +\frac{e}{c}\bs{\nabla}\chi(\bs{r},t),
\end{eqnarray}
and $U\hat{\bs{p}}^2U^\dagger=\hat{\bs{p}}'\bs{\cdot}\hat{\bs{p}}'$.
Denoting the gauge transformed Hamiltonian as $\hat{H}'=U\hat{H}U^\dagger$,
the TDSE after the transformation, $\zI\hbar\,\partial_t'\psi'=\hat{H}'\psi'$, takes
the following form:
\begin{eqnarray}
&&
\zI\hbar\biggl(\frac{\partial}{\partial t}+\frac{\zI e}{\hbar c}\frac{\partial\chi(\bs{r},t)}{\partial t}\biggr)\psi'(\bs{r},t) \nonumber\\[1mm]
&&\hspace{5mm} = \Biggl[ \frac{1}{2m}\biggl(\hat{\bs{p}}+\frac{e}{c}\bs{\nabla}\chi(\bs{r},t)-\frac{e}{c}\bs{A}(\bs{r},t)-\frac{e}{c}\bs{\nabla}\chi(\bs{r},t)\biggr)^2  \nonumber\\
&&\hspace{10mm}
+ e\biggl(\phi(\bs{r},t)-\frac{1}{c}\frac{\partial\chi(\bs{r},t)}{\partial t}\biggr) \Biggl]\psi'(\bs{r},t),
\end{eqnarray}
namely,
\begin{equation}
\zI\hbar\frac{\partial\psi'(\bs{r},t)}{\partial t} = \hat{H}\psi'(\bs{r},t).
\end{equation}
Thus, the transformed wave function $\psi'(\bs{r},t)$ obeys exactly
the same TDSE as $\psi(\bs{r},t)$, differing merely by the phase.
It is therefore possible to arbitrarily choose the function $\chi(\bs{r},t)$
to change the expression.

Here we consider the case where electric field is constant in space
and there is no source of it. It corresponds to the dipole approximation
in atomic physics, which can be expressed by a constant vector potential
$\bs{A}(t)$ and $\phi=0$. In the so-called \textit{length gauge},
$\chi(\bs{r},t)$ is taken to be
\begin{equation}
\chi(\bs{r},t) = -\bs{A}(t)\bs{\cdot}\bs{r}.
\label{Eq:chi-L}
\end{equation}
In this case we have
\begin{eqnarray}
\bs{A}'(t) &=& 0, \\[1mm]
\phi'(\bs{r},t) &=& -\bs{E}(t)\bs{\cdot}\bs{r},
\end{eqnarray}
where $\bs{E}(t)=-\frac{1}{c}\partial\bs{A}(t)/\partial t$ and the TDSE becomes
\begin{equation}
\zI\hbar\frac{\partial\psi^\text{L}(\bs{r},t)}{\partial t}
= \biggl[ \frac{\hat{\bs{p}}^2}{2m} - e\bs{E}(t)\bs{\cdot}\bs{r} \biggr]\psi^\text{L}(\bs{r},t).
\label{Eq:TDSE-L}
\end{equation}
In this representation, the vector potential is absent and a linear scalar potential
$-\bs{E}(t)\bs{\cdot}\bs{r}$ is responsible for the electric field. Since the electric
field couples to the coordinate $\bs{r}$, this choice is called the length gauge.
The superscript `L' here indicates that the wave function is in the length gauge.
While, in the so-called \textit{velocity gauge}, $\chi(\bs{r},t)$ is taken to be
\begin{equation}
\chi(t) = \frac{e}{2mc}\int_0^t\bs{A}^2(t')dt'.
\end{equation}
In this case we have
\begin{eqnarray}
\bs{A}'(t) &=& \bs{A}(t), \\[1mm]
\phi'(t) &=& -\frac{e}{2mc^2}\bs{A}^2(t),
\end{eqnarray}
and the TDSE becomes
\begin{equation}
\zI\hbar\frac{\partial\psi^\text{V}(\bs{r},t)}{\partial t}
= \frac{1}{2m}\biggl[ \hat{\bs{p}}^2 -\frac{2e}{c}\bs{A}(t)\bs{\cdot}\hat{\bs{p}} \biggr]\psi^\text{V}(\bs{r},t).
\label{Eq:TDSE-V}
\end{equation}
Since the vector potential couples to $\hat{\bs{p}}/m$, the velocity in
a classical sense, this choice is called the velocity gauge. The superscript
`V' here indicates that the wave function is in the velocity gauge.

The length and the velocity gauges are related to each other. Suppose
that we are working in the length gauge \eqref{Eq:TDSE-L}, and consider
an inverse transformation of the wave function, i.e. $\psi'(\bs{r},t)=
U^\dagger(\bs{r},t)\psi^\text{L}(\bs{r},t)$. It leads to
\begin{equation}
\zI\hbar\frac{\partial\psi'(\bs{r},t)}{\partial t} = \frac{1}{2m}\biggl[ \hat{\bs{p}}^2-\frac{2e}{c}\bs{A}(t)\bs{\cdot}\hat{\bs{p}}+\frac{e^2}{c^2}\bs{A}^2(t)\biggr]\psi'(\bs{r},t).
\end{equation}
Since $\bs{A}^2(t)$ is constant in space and the last term in the squared
parentheses affects only the phase of the wave function, this representation
can also be regarded as the velocity gauge [cf.\ Eq.~\eqref{Eq:TDSE-V}].
In Sec.~\ref{Sec:TDDFT}, actually we first consider TDKS equations in the
length gauge, Eq.~\eqref{TDKS-l}, where a linear scalar potential, $\phi(\bs{r},t)
=-\bs{E}(t)\bs{\cdot}\bs{r}$, is introduced to express a constant external
electric field, and then an inverse gauge transformation is performed to
eliminate the scalar potential. This explains the minus sign in the exponent
in Eq.~\eqref{Eq:u_tilde}, which corresponds to the inverse transformation
with Eqs.~\eqref{Eq:Utrans} and \eqref{Eq:chi-L}.


\begin{thebibliography}{99}
\bibitem{Ravenhall(1983)}
D.G.~Ravenhall, C.J.~Pethick, and J.R.~Wilson,
Structure of Matter below Nuclear Saturation Density,
Phys. Rev. Lett. \textbf{50}, 2066 (1983).
\bibitem{Hashimoto(1984)}
M.~Hashimoto, H.~Seki, and M.~Yamada, Shape of Nuclei in the Crust of Neutron Star,
Prog. Theor. Phys. \textbf{71}, 320 (1984).
\bibitem{Chamel(2017)}
N.~Chamel, Entrainment in Superfluid Neutron-Star Crusts: Hydrodynamic Description and Microscopic Origin,
J. Low Temp. Phys. \textbf{189}, 328 (2017).
\bibitem{Ashcroft-Mermin}
N.W.~Ashcroft, N.D.~Mermin, and D.~Wei, \textit{Solid State Physics Revised Edition}
(Cengage Lerning Asia Pte Ltd, 2016).

\bibitem{Oyamatsu(1994)}
K.~Oyamatsu and Y.~Yamada, Shell energies of non-spherical nuclei in the inner crust of a neutron star,
Nucl. Phys. \textbf{A578}, 184 (1994).
\bibitem{Carter(2005)}
B.~Carter, N.~Chamel, and P.~Haensel,
Entrainment coefficient and effective mass for conduction neutrons in neutron star crust: simple microscopic models,
Nucl. Phys. \textbf{A748}, 675 (2005).
\bibitem{Chamel(2005)}
N.~Chamel, Band structure effects for dripped neutrons in neutron star crust,
Nucl. Phys. \textbf{A747}, 109 (2005).
\bibitem{Chamel(2006)}
N.~Chamel, Effective mass of free neutrons in neutron star crust,
Nucl. Phys. \textbf{A773}, 263 (2006).
\bibitem{Chamel(2007)}
N.~Chamel, S.~Naimi, E.~Khan, and J.~Margueron,
Validity of the Wigner-Seitz approximation in neutron star crust,
Phys. Rev. C \textbf{75}, 055806 (2007).
\bibitem{Chamel(2012)}
N.~Chamel, Neutron conduction in the inner crust of a neutron star in the framework of the band theory of solids,
Phys. Rev. C \textbf{85}, 035801 (2012).

\bibitem{Chamel(2013)}
N.~Chamel, D.~Page, and S.~Reddy,
Low-energy collective excitations in the neutron star inner crust,
Phys. Rev. C \textbf{87}, 035803 (2013).
\bibitem{Kobyakov(2013)}
K.~Kobyakov and C.J.~Pethick,
Dynamics of the inner crust of neutron stars: Hydrodynamics, elasticity, and collective modes,
Phys. Rev. C \textbf{87}, 055803 (2013).
\bibitem{Kobyakov(2016)}
K.~Kobyakov and C.J.~Pethick,
Nucleus-nucleus interactions in the inner crust of neutron stars,
Phys. Rev. C \textbf{94}, 055806 (2016).
\bibitem{Durel(2018)}
D.~Durel and M.~Urban, Long-wavelength phonons in the crystalline and pasta phases of neutron-star crusts,
Phys. Rev. C \textbf{97}, 065805 (2018).

\bibitem{Anderson(1975)}
P.W.~Anderson and N.~Itoh, Pulsar glitches and restlessness as a hard superfluidity phenomenon,
Nature \textbf{256}, 25 (1975). 

\bibitem{Andersson(2012)}
N.~Andersson, K.~Glampedakis, W.C.G.~Ho, and C.M.~Espinoza,
Pulsar Glitches: The Crust is not Enough,
Phys. Rev. Lett. \textbf{109}, 241103 (2012).
\bibitem{Chamel(2013)glitch}
N.~Chamel, Crustal Entrainment and Pulsar Glitches,
Phys. Rev. Lett. \textbf{110}, 011101 (2013).
\bibitem{Haskell(2015)}
B.~Haskell and A.~Melatos, Models of pulsar glitches,
Int. J. Mod. Phys. D \textbf{24}, 153008 (2015).

\bibitem{Watanabe(2017)}
G.~Watanabe and C.J.~Pethick,
Superfluid Density of Neutrons in the Inner Crust of Neutron Stars: New Life for Pulsar Glitch Models,
Phys. Rev. Lett. \textbf{119}, 062701 (2017).
\bibitem{Sauls(2020)}
J.A.~Sauls, N.~Chamel, and M.A.~Alpar, Superfluidity in Disordered Neutron Star Crust,
arXiv:2001.09959 [astro-ph.HE].

\bibitem{Martin(2016)}
N.~Martin and M.~Urban, Superfluid hydrodynamics in the inner crust of neutron stars,
Phys. Rev. C \textbf{94}, 065801 (2016).

\bibitem{Kashiwaba(2019)}
Yu Kashiwaba and T.~Nakatsukasa, Self-consistent band calculations of the slab phase in the neutron-star crust,
Phys. Rev. C \textbf{100}, 035804 (2019).

\bibitem{Negele(review)}
J.W.~Negele, The mean-field theory of nuclear structure and dynamics,
Rev. Mod. Phys. \textbf{54}, 913 (1982).
\bibitem{BKN(1976)}
P.~Bonche, S.~Koonin, and J.W.~Negele,
One-dimensional nuclear dynamics in the time-dependent Hartree-Fock approximation,
Phys. Rev. C \textbf{13}, 1226 (1976).
\bibitem{Nakatsukasa(PTEP)}
T.~Nakatsukasa, Density functional approaches to collective phenomena in nuclei: Time-dependent density functional theory for perturbative and non-perturbative nuclear dynamics,
Prog. Theor. Exp. Phys. \textbf{2012}, 01A207 (2012).
\bibitem{Simenel(review)}
C.~Simenel, Nuclear quantum many-body dynamics,
Eur. Phys. J. A \textbf{48}, 152 (2012).
\bibitem{Nakatsukasa(review)}
T.~Nakatsukasa, K.~Matsuyanagi, M.~Matsuo, and K.~Yabana, Time-dependent density-functional description of nuclear dynamics,
Rev. Mod. Phys. \textbf{88}, 045004 (2016).
\bibitem{TDHF-review(2018)}
C.~Simenel and A.S.~Umar, Heavy-ion collisions and fission dynamics with the time-dependent Hartree-Fock theory and its extensions,
Prog. Part. Nucl. Phys. \textbf{103}, 19 (2018).
\bibitem{Stevenson(2019)}
P.D.~Stevenson and M.C.~Barton, Low-energy heavy-ion reactions and the Skyrme effective interaction,
Prog. Part. Nucl. Phys. \textbf{104}, 142 (2019).
\bibitem{Sekizawa(2019)}
K.~Sekizawa, TDHF Theory and Its Extensions for Multinucleon Transfer Reactions: A Mini Review,
Front. Phys. \textbf{7}, 20 (2019).

\bibitem{DFT1}
P.~Hohenberg and W.~Kohn, Inhomogeneous Electron Gas, Phys. Rev. \textbf{136}, B864 (1964).
\bibitem{DFT2}
W.~Kohn and L.J.~Sham, Self-Consistent Equations Including Exchange and Correlation Effects,
Phys. Rev. \textbf{140}, A1133 (1965).
\bibitem{DFT3}
E.~Runge and E.K.U.~Gross, Density-Functional Theory for Time-Dependent Systems,
Phys. Rev. Lett. \textbf{52} (1984) 997.
\bibitem{DFT4}
W.~Kohn, Nobel Lecture: Electronic structure of matter---wave functions and density functionals,
Rev. Mod. Phys. \textbf{71}, 1253 (1999).

\bibitem{Yabana(1996)}
K.~Yabana and G.F.~Bertsch,
Time-dependent local-density approximation in real time,
Phys. Rev. B \textbf{54}, 4484 (1996).
\bibitem{Bertsch(2000)}
G.F.~Bertsch, J.-I.~Iwata, A.~Rubio, and K.~Yabana,
Real-space, real-time method for the dielectric function,
Phys. Rev. B \textbf{62}, 7998 (2000).

\bibitem{Octopus}
X.~Andrade, D.~Strubbe, U.~De Giovannini, A.~Hjorth Larsen, M.J.T.~Oliveira, J.~Alberdi-Rodriguez,
A.~Varas, I.~Theophilou, N.~Helbig, M.J.~Verstraete, L.~Stella, F.~Nogueira, A.~Aspuru-Guzik,
A.~Castro, M.A.L.~Marques, and A.~Rubio,
Real-space grids and the Octopus code as tools for the development of new simulation approaches for electronic systems,
Phys. Chem. Chem. Phys. \textbf{17}, 31371 (2015); \url{https://octopus-code.org/wiki/Main_Page}.
\bibitem{SALMON}
M.~Noda, S.A.~Sato, Y.~Hirokawa, M.~Uemoto, T.~Takeuchi, S.~Yamada, A.~Yamada, Y.~Shinohara,
M.~Yamaguchi, K.~Iida, I.~Floss, T.~Otobe, K.-M.~Lee, K.~Ishimura, T.~Boku, G.F.~Bertsch, K.~Nobusada, and K.~Yabana,
SALMON: Scalable Ab-initio Light-Matter simulator for Optics and Nanoscience,
Compt. Phys. Commun. \textbf{235}, 356 (2019); \url{https://salmon-tddft.jp/}.

\bibitem{Yabana(2012)}
K.~Yabana, T.~Sugiyama, Y.~Shinohara, T.~Otobe, and G.F.~Bertsch,
Time-dependent density functional theory for strong electromagnetic fields in crystalline solids,
Phys. Rev. B \textbf{85}, 045134 (2012).
\bibitem{Yamada(2019)}
A.~Yamada and K.~Yabana,
Multiscale time-dependent density functional theory for a unified description of ultrafast dynamics: Pulsed light, electron, and lattice motions in crystalline solids,
Phys. Rev. B \textbf{99}, 245103 (2019).

\bibitem{Anderson(1967)}
P.W.~Anderson, Infrared Catastrophe in Fermi Gases with Local Scattering Potentials,
Phys. Rev. Lett. \textbf{18}, 1049 (1967).

\bibitem{Chabanat(1998)}
E.~Chabanat, P.~Bonche, P.~Haensel, J.~Meyer, and R.~Schaeffer,
A Skyrme parametrization from subnuclear to neutron star densities Part II. Nuclei far from stabilities,
Nucl. Phys. \textbf{A635}, 231 (1998); \textbf{A643}, 441 (1998).
\bibitem{Bender(2003)}
M.~Bender, P.-H.~Heenen, and P.-G.~Reinhard,
Self-consistent mean-field models for nuclear structure,
Rev. Mod. Phys. \textbf{75}, 121 (2003).
\bibitem{Lesinski(2007)}
T.~Lesinski, M.~Bender, K.~Bennaceur, T.~Duguet, and J.~Meyer,
Tensor part of the Skyrme energy density functional: Spherical nuclei,
Phys. Rev. C \textbf{76}, 014312 (2007).
\bibitem{Kortelainen(2010)}
M.~Kortelainen, R.J.~Furnstahl, W.~Nazarewicz, and M.V.~Stoitsov,
Natural units for nuclear energy density functional theory,
Phys. Rev. C \textbf{82}, 011304(R) (2010).

\bibitem{Baldo(2013)}
M.~Baldo, L.M.~Robledo, P.~Schuck, and X.~Vi\~nas,
New Kohn-Sham density functional based on microscopic nuclear and neutron matter equations of state,
Phys. Rev. C \textbf{87}, 064305 (2013).
\bibitem{Baldo(2017)}
M.~Baldo, L.M.~Robledo, P.~Schuck, and X.~Vi\~nas,
Barcelona-Catania-Paris-Madrid functional with a realistic effective mass,
Phys. Rev. C \textbf{95}, 014318 (2017).

\bibitem{Yabana(2006)}
K.~Yabana, T.~Nakatsukasa, J.-I.~Iwata, and G.F.~Bertsch,
Real-time, real-space implementation of the linear response time-dependent density-functional theory,
Phys. Stat. Sol. B \textbf{243}, 1121 (2006).

\bibitem{Papakonstantinou(2013)}
P.~Papakonstantinou, J.~Margueron, F.~Gulminelli, and Ad.R.~Raduta,
Densities and energies of nuclei in dilute matter at zero temperature,
Phys. Rev. C \textbf{88}, 045805 (2013).

\bibitem{Inakura(2017)}
T.~Inakura and M.~Matsuo, Anderson-Bogoliubov phonons in the inner crust of neutron stars:
Dipole excitation in a spherical Wigner-Seitz cell,
Phys. Rev. C \textbf{96}, 025806 (2017).
\bibitem{Inakura(2019)}
T.~Inakura and M.~Matsuo, Coexistence of Anderson-Bogoliubov phonon and quadrupole
vibration in the inner crust of neutron stars,
Phys. Rev. C \textbf{99}, 045801 (2019).

\bibitem{Engel(1975)}
Y.M.~Engel, D.M.~Brink, K.~Goeke, S.J.~Krieger, and D.~Vautherin,
Time-dependent Hartree-Fock theory with Skyrme's interaction,
Nucl. Phys. \textbf{A249}, 215 (1975).
\bibitem{Chamel(2019)}
N.~Chamel and V.~Allard,
Entrainment effects in neutron-proton mixtures within the nuclear energy-density functional theory: Low-temperature limit,
Phys. Rev. C \textbf{100}, 065801 (2019).

\bibitem{Bulgac(2012)}
A.~Bulgac, P.~Magierski, and M.M.~Forbes,
\textit{The Unitary Fermi Gas: From Monte Carlo to Density Functionals},
in: BCS-BEC Crossover and the Unitary Fermi Gas (ed.) W. Zwerger;
Lecture Notes in Physics, Vol. \textbf{836}, pp. 305--373 (Springer, Heidelberg, 2012).
\bibitem{Bulgac(2019)}
A.~Bulgac, Time-Dependent Density Functional Theory for Fermionic Superfluids: From Cold Atomic Gases--To Nuclei and Neutron Stars Crust, Phys. Status Solidi B \textbf{256}, 1800592 (2019).

\bibitem{Daniel(2021)}
D.~P\k{e}cak, N.~Chamel, P.~Magierski, and G.~Wlaz{\l}owski,
Properties of a quantum vortex in neutron matter at finite temperatures,
Phys. Rev. C \textbf{104}, 055801 (2021).

\bibitem{Schuetrumpf(2013)}
B.~Schuetrumpf, M.A.~Klatt, K.~Iida, J.A.~Maruhn, K.~Mecke, and P.-G.~Reinhard,
Time-dependent Hartree-Fock approach to nuclear ``pasta'' at finite temperature,
Phys. Rev. C \textbf{87}, 055805 (2013).
\bibitem{Schuetrumpf(2014)}
B.~Schuetrumpf, K.~Iida, J.A.~Maruhn, and P.-G.~Reinhard,
Nuclear ``pasta matter'' for different proton fractions,
Phys. Rev. C \textbf{90}, 055802 (2014).
\bibitem{Schuetrumpf(2015)1}
B.~Schuetrumpf, M.A.~Klatt, K.~Iida, G.E.~Schr\"oder-Turk, J.A.~Maruhn, K.~Mecke, and P.-G.~Reinhard,
Appearance of the single gyroid network phase in ``nuclear pasta'' matter,
Phys. Rev. C \textbf{91}, 025801 (2015).
\bibitem{Schuetrumpf(2015)2}
B.~Schuetrumpf and W.~Nazarewicz,
Twist-averaged boundary conditions for nuclear pasta Hartree-Fock calculations,
Phys. Rev. C \textbf{92}, 045806 (2015).

\bibitem{VNdyn(2016)}
G.~Wlaz\l{}owski, K.~Sekizawa, P.~Magierski, A.~Bulgac, and M.M.~Forbes,
Vortex Pinning and Dynamics in the Neutron Star Crust,
Phys. Rev. Lett. \textbf{117}, 232701 (2016).

\bibitem{Miller(2019)}
M.C.~Miller, C.~Chirenti, and T.E.~Strohmayer,
On the Persistence of QPOs during the SGR 1806-20 Giant Flare,
Astro. Phys. J \textbf{871}, 95 (2019).
\bibitem{Sotani(2019)}
H.~Sotani, K.~Iida, and K.~Oyamatsu,
Astrophysical implications of double-layer torsional oscillations in a neutron star crust as a lasagna sandwich,
MNRAS \textbf{489}, 3022 (2019).
\end{thebibliography}
\end{document}